\documentclass[lettersize,journal]{IEEEtran}
\usepackage{amsmath,amsfonts}
\usepackage{algorithmic}
\usepackage{algorithm}
\usepackage{array}
\usepackage{textcomp}
\usepackage{stfloats}
\usepackage{url}
\usepackage{verbatim}
\usepackage{graphicx}
\usepackage{subcaption}
\usepackage{cite}
\usepackage{amssymb}
\usepackage{bm}

\usepackage[all]{nowidow}
\usepackage{pgfplots}
\pgfplotsset{compat=newest}
 \usepackage{tabularx}
\usepackage{booktabs}
\usepackage{adjustbox}
\usepackage{gensymb}
\usepackage[bookmarks=false]{hyperref}
\usepackage{makecell}
\usepackage{glossaries}
\usepackage{orcidlink}
\usepackage{siunitx}
\newacronym{avp}{AVP}{Automated Valet Parking}
\newacronym{cfar}{CFAR}{Constant False Alarm Rate}
\newacronym{cnn}{CNN}{Convolutional Neural Network}
\newacronym{dbscan}{DBSCAN}{Density-Based Spatial Clustering of Applications with Noise}
\newacronym{dsp}{DSP}{Digital Signal Processing}
\newacronym{hdbscan}{HDBSCAN}{Hierarchical Density-Based Spatial Clustering of Applications with Noise}
\newacronym{ekf}{EKF}{Extended Kalman Filter}
\newacronym{fit}{FIT}{Failure-in-time}
\newacronym{odd}{ODD}{Operational Design Domain}
\newacronym{ogm}{OGM}{Occupancy Grid Map}
\newacronym{rcs}{RCS}{Radar Cross Section}
\newacronym{rtk}{RTK}{Real Time Kinematic}
\newacronym{sotif}{SOTIF}{Safety of the Intended Functionality}
\newacronym{snr}{SNR}{Signal-to-noise ratio}
\newacronym{ttc}{TTC}{Time-to-Collision}
\newacronym{vru}{VRU}{Vulnerable Road User}
\newacronym{vrus}{VRUs}{Vulnerable Road Users}

\begin{document}

\title{SAFERad: A Framework to Enable Radar Data for Safety-Relevant Perception Tasks}

\author{Tim Brühl \orcidlink{0009-0006-9322-2850}, Jenny Glönkler, Robin Schwager, Tin Stribor Sohn, \\ Tim Dieter Eberhardt, and Sören Hohmann, \textit{Senior Member, IEEE}
    % <-this % stops a space
    \thanks{Manuscript received November 18, 2024; revised December 31, 2024. (\textit{Corresponding author: Tim Brühl.})}
    \thanks{This work has been submitted to the IEEE for possible publication. Copyright may be transferred without notice, after which this version may no longer be accessible.}
    \thanks{Tim Brühl is with Dr. Ing. h.c. F. Porsche AG, Weissach, Germany, and with Karlsruhe Institute of Technology, Germany. (e-mail: \href{mailto:tim.bruehl@kit.edu}{tim.bruehl@kit.edu})}
    \thanks{Jenny Glönkler is with Dr. Ing. h.c. F. Porsche AG, Weissach, Germany, and with Esslingen University of Applied Sciences, Germany. (e-mail: \href{mailto:jenny.gloenkler@web.de}{jenny.gloenkler@web.de})}
    \thanks{Robin Schwager, Tin Stribor Sohn and Tim Dieter Eberhardt are with Dr. Ing. h.c. F. Porsche AG, Weissach, Germany. (e-mail: \{robin.schwager1, tin\_stribor.sohn, tim.eberhardt1\}@porsche.de)}
    \thanks{Sören Hohmann is with Karlsruhe Institute of Technology, Germany. (e-mail: \href{mailto:soeren.hohmann@kit.edu}{soeren.hohmann@kit.edu})}}

\maketitle

\begin{abstract}
    % 150-250 words
    Radar sensors play a crucial role for perception systems in automated driving but suffer from a high level of noise. In the past, this could be solved by strict filters, which remove most false positives at the expense of undetected objects. Future highly automated functions are much more demanding with respect to error rate. Hence, if the radar sensor serves as a component of perception systems for such functions, a simple filter strategy cannot be applied. In this paper, we present a modified filtering approach which is characterized by the idea to vary the filtering depending on the potential of harmful collision with the object which is potentially represented by the radar point. We propose an algorithm which determines a criticality score for each point based on the planned or presumable trajectory of the automated vehicle. Points identified as very critical can trigger manifold actions to confirm or deny object presence. Our pipeline introduces criticality regions. The filter threshold in these criticality regions is omitted. Commonly known radar data sets do not or barely feature critical scenes. Thus, we present an approach to evaluate our framework by adapting the planned trajectory towards vulnerable road users, which serve as ground truth critical points. Evaluation of the criticality metric prove high recall rates. Besides, our post-processing algorithm lowers the rate of non-clustered critical points by \SI{74.8}{\percent} in an exemplary setup compared to a moderate, generic filter.

\end{abstract}

\begin{IEEEkeywords}
    Radar object detection, safety-aware processing, criticality assessment, noise filtering
\end{IEEEkeywords}

\section{INTRODUCTION}
\label{sec:Introduction}
Automated driving and parking functions are fields which currently receive a high attention in research. Here, a major demand on automated vehicles is a safe behavior in all situations of the \gls{odd}. In SAE level 2 functions, it is possible to handle critical edge cases by the driver who is, in case of uncertainty, still in charge of the vehicle’s action. This is a fact which is only partly valid for SAE level 3 functions and not valid for SAE level 4 functions anymore since the driver is completely detached from the vehicle \cite{SAE2021}. The requirements on \gls{sotif} are incomparably higher in this case. Guaranteeing this safety is a process which stretches across the perception, planning, and actor control tasks which are all involved in the function. However, the downstream elements can only act safely if the perception element works soundly.

Perception is nowadays performed by a combination of manifold sensors inside of a sensor set \cite{Yeong2021}. While lidar and camera sensors suffer from heavy performance degradation in bad weather conditions, radar technologies only need to put up with minor limitations in range. They currently seem to be the technology which gives reliance in situations where camera and lidar fail \cite{Zhang2023}. However, attributable to their physical measurement principle, radar sensors come with other weaknesses. One drawback is the poor sensitivity to objects with a low \gls{rcs}, e.g., pedestrians or objects with an absorbing or permeable material. The \gls{rcs} measures the reflectivity of an object, more precisely, how much input power is reflected by an object \cite{Knott1988}. Another disadvantage is a commonly sparse and noisy point cloud. For this reason, the numerous noise points are filtered to reduce the number of false positives. The objective is to reach a solid balance between a high detection rate and an acceptably low probability for False Positives. One popular filter criterion might be the \gls{rcs} value, since weak reflections tend to be False Positive detections which do not belong to an existing object. The downside of this procedure is that the erasure of several true positive points is accepted, which may result in a descent of functional safety. This is not a viable approach if a safe radar perception with an upper limit of errors per hour shall be guaranteed. Error prevention can take place in two different ways: The first one is ensuring the detectability of objects with minor \gls{rcs} by improving the general \gls{snr}. This can be achieved by development of the sensor properties, antenna technology, and electrical components in the radar's hardware. The second one is to ensure that every true positive point whose disregard may lead to injuries during the operation of the automated function is considered in the perception algorithm.

The problem of current approaches is that they concede undetected objects for a more robust overall functionality. In their filtering methods, the potential impact of measurements on the driving task, subsequently called the \textit{criticality} of the radar point, is not respected. In this paper, we tackle this drawback by altering the filter algorithm in dependence of the point's criticality. For this purpose, we present a pipeline consisting of two steps:

\begin{enumerate}
    \item Potentially critical points are identified by considering the current motion state of the vehicle and the prospective intentions of the automated functions.
    \item The critical points are processed in a way differing from processing non-critical points. Optionally, information from past measurements or image information can be included to further enrich critical points with information.
\end{enumerate}

To sum it up, these are the main contributions of this paper:
\begin{itemize}
    \item An algorithm to evaluate criticality for radar detections on a raw point level
    \item A pipeline to treat critical points by filter threshold adaption
    \item An experimental study performed on an open data set which evaluates the performance of the criticality evaluation and point treatment
\end{itemize}

The structure of the paper is as follows: in section \ref{sec:related_work}, an overview about state-of-the-art point processing algorithms, criticality-aware information processing, and approaches of sensor fusion is given - specifically in the application field of automated driving. In section \ref{sec:method}, an approach for the evaluation of critical points inside of the vehicle’s driving tube is introduced. Additionally, the post-processing by lowering the signal power threshold is outlined. Section \ref{sec:experimental_setup} starts with a description of the adaptions to the used data set to make them applicable for the safety investigations. Next, the data set as well as the setup applied for the real-world tests is depicted. The following section \ref{sec:experimental_results} gives an overview about the experimental results. Section \ref{sec:discussion} discusses the outcomes and limitations of this research before section \ref{sec:conclusion} concludes and points to fields for further research.

%At an early stage, a common strategy is to apply the \gls{cfar} thresholding algorithm to extract objects from the continuous signal spectrum \cite{Rohling1983, Zebiri2021}. The \gls{cfar} algorithm finds detections in the power spectrum by calculating a threshold as a multiple of the estimated noise floor. Detections are exceedings of this threshold by the power spectrum. Peaks of small objects, which protude not enough from this noise floor, can unintentionally be neglected.

\section{RELATED WORK}
\label{sec:related_work}

In this section, the state of the art in radar signal processing and object detection under safety aspects is delineated. We contrast our ideas to the ones which are currently described in literature.

Forming objects is one important step in radar signal processing. Srivastav and Mandal partition this step into the three substages detection, association, and tracking, which must be passed through in both early and late fusion approaches \cite{Srivastav2023}. In detail, the clustering and tracking takes part after static noise is removed. As an established clustering technique, \gls{dbscan} is named, which can be improved by several variants \cite{Pearce2023}. The native \gls{dbscan} clustering approach determines clusters based on their density. Core objects, which have a certain number $min_{pts}$ of neighbors within the distance $\epsilon$, are connected to clusters if all core objects are neighbors of each other. The neighbors of each core point, which can either be other core points or border points, belong to the formed cluster. Thus, the shapes of clusters can be arbitrary, and the algorithm relies only on the two parameters $\epsilon$ and $min_{pts}$, where the paper suggests setting $min_{pts} = 4$ due to investigations of the k-dist graph \cite{ester1996density}. We apply native \gls{dbscan} clustering as it groups points meaningful. We address the weakness of suppressing single reflection points by distinguishing between clustering problems and flaws in our method in section \ref{sec:experimental_results}.

Research evolved the hypothesis that data processing requires to focus rather on a guaranteed detection and computing performance in critical areas than a maximization of the detection rate. There is sizable research about the definition of these areas.\cite{Wolf2021} apply \gls{ttc} and distance to the object as metrics for criticality. \cite{Topan2023} and \cite{schonemann2019maneuver} define perception safety zones based on the current objective of the vehicle. Our approach is geared to these concepts but defines criticality dependent on the trajectory of the automated function. Furthermore, it calculates criticality per radar point, focusing on the area which is intended to run on by the automated function. This is diverging from the related approaches that determine criticality per object.

In the context of image and lidar processing, it is pointed out that prioritized queues, selected from criticality assignment and attention cueing, should be processed first. Processed bounding boxes with a low \gls{ttc} inside of the observer’s trajectory are prioritized. Candidate regions in images can be identified based on, e.g., optical flow and receive higher processing priority and fidelity \cite{Liu2023}. Other works solely focus on the identification of critical areas inside images. Lyssenko et al. \cite{Lyssenko2022} suggest to include information about the trajectory of the vehicle and the surrounding pedestrians in image object detectors such as RetinaNet \cite{Lin_2017_ICCV} which is used in their experiments. Philipp et al. \cite{Philipp2022} examine the error tolerance for object detectors by evaluating the criticality of objects in a simplified set of scenarios, so-called basic areas, using an own approach which is comparable to the one of \cite{schonemann2019maneuver}. Another work determines the p-quantile relative scenario risk which is the prevalent risk for certain objects being not detected in a driving scene \cite{antonante2023task}. While these approaches work on a bounding box level and propose to prioritize the processing of critical objects, we claim that starting the processing of critical points in an earlier stage, e.g., the point stage, and to rethink all processing steps which can lead to loss of potentially relevant information, is beneficial.

Recently, Ceccarelli and Montecchi \cite{Ceccarelli2024} highlight that objects should be considered from both an existence confidence and a safety point of view. Three hypotheses were stated: first, removing uninteresting objects will be beneficial for the object detector's performance. Second, objects with high criticality should always be included in the planning task. Third, an optimal decision can be made to filter an object based on the two factors confidence and criticality. The work does not investigate the actual collision probability. Secondly, comparable with the other works and contrary to our idea, the filtering took part in a late step, after the object detection task. Work on point clouds is sparser, however, \cite{Feng2024} introduce PolarPoint-BEV, a bird's-eye-view representation of a point cloud which is reduced by a network with explainability focus for only criticality-relevant points.

To conclude, the related work lacks a method which interprets safety-aware signal processing on an early level. To avoid false negative objects, adapted approaches must be applied in all signal processing steps. Our work tries to understand the root of hazardous situations, which can lead to injured road participants. Contrary to the prevalent work, it traces back these situations as far as possible within the perception process.

% Campello et al. name different drawbacks of native \gls{DBSCAN} algorithms, e.g., that data is labeled without hierarchy, the poor interpretability of hierarchical clustering algorithms and a high number of input parameters for the algorithms. As a result, they propose \gls{HDBSCAN}, which gets along with a single algorithm $m_{pts}$ which controls the density of the formed clusters. The value $\lambda$ which is the reciprocal of the parameter $\epsilon$ known from the \gls{DBSCAN} algorithm is chosen automatically based on the excess of mass criterion, which balances between cluster size and cluster density. One advantage of \gls{HDBSCAN} is that prior knowledge about points belonging to one group can be included during the clustering algorithm \cite{Density-Based Clustering Based on Hierarchical Density Estimates}.

\section{METHOD}
\label{sec:method}

In this section, the main ideas for a safety-aware point filtering pipeline are elucidated in detail. The main elements are the criticality point evaluation, considering the planned actions of the vehicle, and the post-processing of radar point cloud, adding points to the pipeline which have previously been filtered.

\subsection{Framework overview}
\label{sec:framework}

\begin{figure}[ht]
	\centering
	\includegraphics[width = \linewidth]{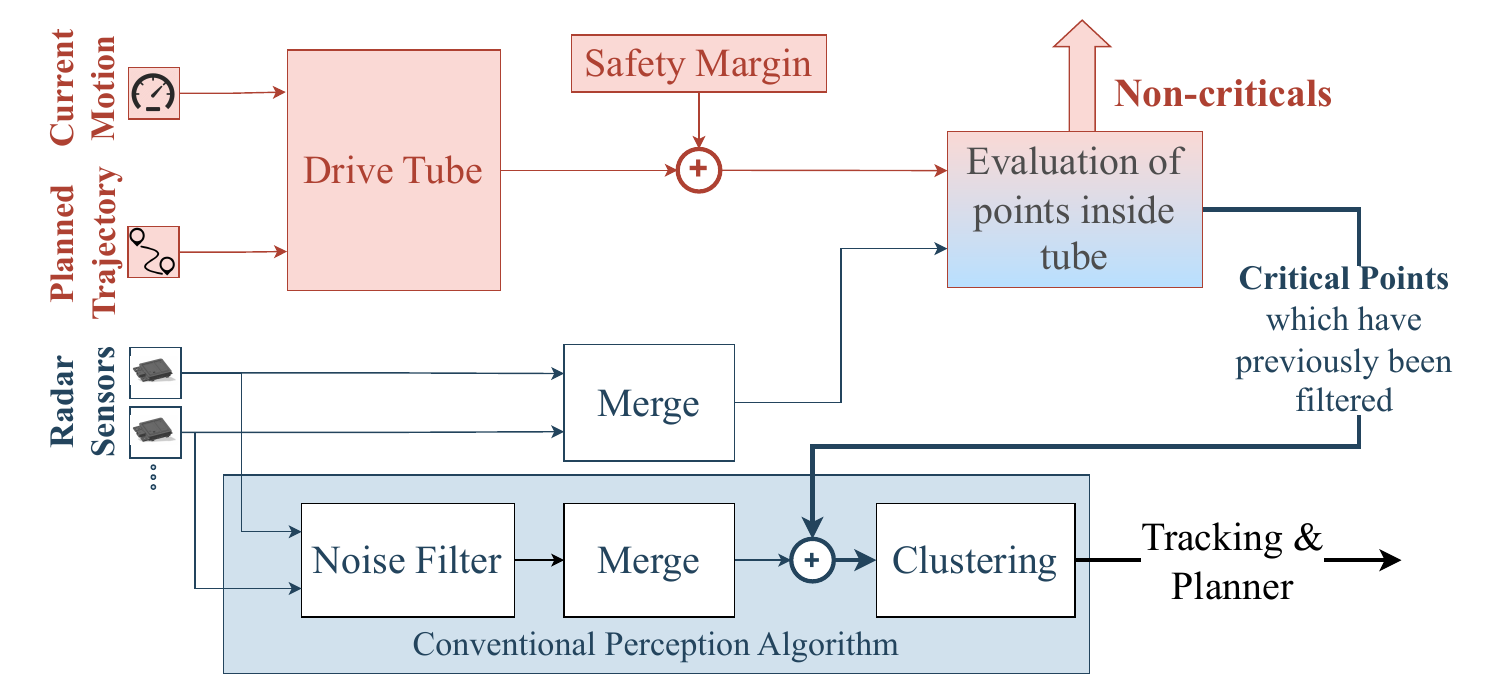}
	\caption{Proposal for a concept for safety-aware radar point processing in perception tasks}
	\label{fig:framework}
\end{figure}

The main concept with all of the single elements, which will be explained during the next sections, is shown in figure \ref{fig:framework}.

It consists of a conventional perception pipeline, which is depicted in the blue box in the lower part of the figure, as well as a criticality estimation for each point which is based on the current driving tube, highlighted by the red boxes. Non-critical points whose existence seems unlikely will be deleted, while the filter thresholds for points in critical regions are removed. By this means, points with a weak signal are fed to the conventional pipeline during a feedback loop. This allows for utilizing prevalent evidence of object existence which is delivered by the raw radar point cloud.

\subsection{Drive-tube based evaluation of criticality}
\label{sec:drivetube}

\begin{figure}[ht]
	\centering
	\includegraphics[width = \linewidth]{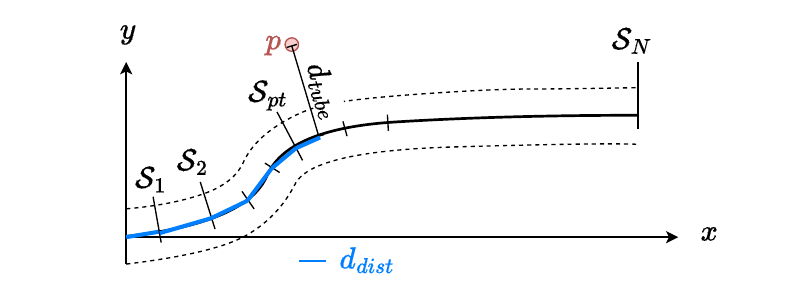}
	\caption{Concept of a discrete driving tube with states $\mathcal{S}$ and the distance $d_{tube}$ to the driving tube}
	\label{fig:drivetube}
\end{figure}

The core idea of the criticality evaluation is that the equal treatment of radar points independent from their impact on the downstream automated function does not seem reasonable. A point which is outside of the action field of the automated function is not critical in the sense that the automated function threatens the integrity of the \gls{vru} represented by this point. Whereas a point inside the driving tube or intended driving tube needs to be treated meticulously to prevent the function of harming potential \gls{vru} represented by the point. In the case of manual operation, the driving tube can be determined from the current velocity and the steering action of the driver. During automated operation, the planned path $\mathcal{P}$ of the automated function is applicable. This path is described as a polygon consisting of $N=15$ discrete states: $\mathcal{P} = \{ \mathcal{S}_{1}, \mathcal{S}_{2}, \dots, \mathcal{S}_{N}\}$. Each state $\mathcal{S}$ features a position $(x, y)$, a driven velocity $v$, the applied acceleration or deceleration $a$, a required steering angle $\psi$, and a yaw angle $\alpha$: $\mathcal{S} = [x, y, v, a, \psi, \alpha]^\intercal$. The horizon of the trajectory is $t = \SI{3.0}{\second}$, meaning that each state $\mathcal{S}_{k}$ has a time shift of $\Delta t = \SI{0.2}{\second}$ from the next one $\mathcal{S}_{k+1}$. The criticality value derives primarily from the point position in the driving tube and the presumable velocity which depends on the current strategy of the planner. These values are determined by finding the perpendicular between the path and the point $p$, which has the distance $d_{tube}$ from the center of the driving path line. The travel distance $d_{dist}$ from the vehicle front to the point is calculated by the summation of the linear approximation of all segments up to the junction with the perpendicular. As the presumable velocity, the state velocity $v_{pt}$ of the state $\mathcal{S}_{pt}$ which is closest to the radar point is applied. In figure \ref{fig:drivetube}, this concept is depicted.

\begin{figure*}[ht]
	\centering
	\includegraphics[width = 2.0\columnwidth]{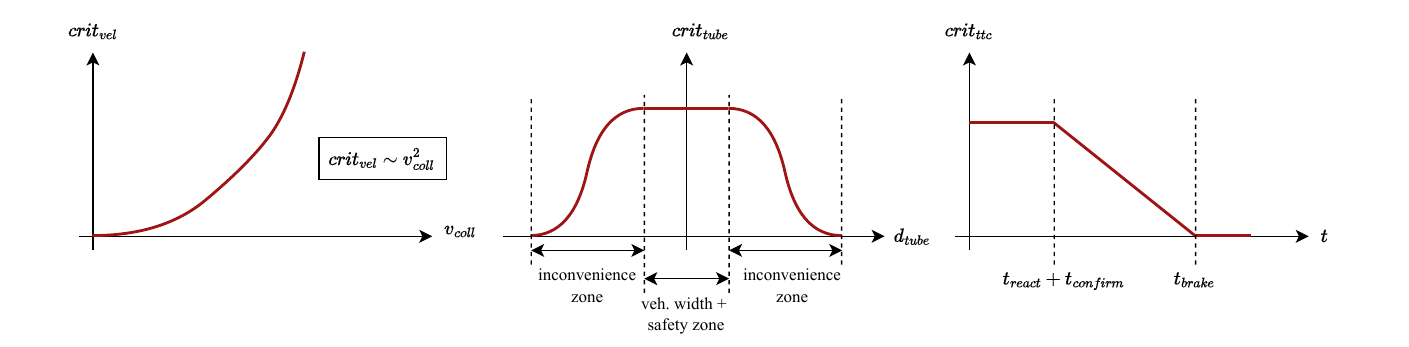}
	\caption{Qualitative figures for the criticality components}
	\label{fig:crit_components}
\end{figure*}

With the parameters which have just been proposed, the criticality score is calculated. Factors which increase the criticality are a high potential collision velocity and a short time for the prospective observation of the point due to lower chances for validation and less reaction time for a braking action. The first factor is represented by the \textit{velocity criticality}, while the \textit{distance criticality} is a measure of the second factor. To assign a measure to the radar point distance criticality, an energy-driven approach is used. This means that a point receives higher distance criticality, the more collision energy ($W = \frac{1}{2} m_{veh} \Delta v^2$) is added by the vehicle which is controlled by the automated driving function in case of a collision. By this means, criticality grows quadratically with the velocity difference $\Delta v$ between the automated vehicle and the object in case of a potential collision. Besides, we introduced the \textit{tube criticality} to measure if the lateral position of the point is inside or near the drive tube. The coherence of parameters and the three independent criticality components - velocity criticality $crit_{vel}$, tube criticality $crit_{tube}$, and distance criticality $crit_{ttc}$ - that form the point criticality $crit_{p}$ is depicted in figure \ref{fig:crit_components}.

In the following, we explain how each criticality component can be determined.

\begin{itemize}

	\item Velocity Criticality $crit_{vel}$: The severity of injury of a \gls{vru} after a potential collision with the vehicle controlled by the automated function depends significantly on the collision speed $v_{coll}$. The vehicle of the mass $m_{veh}$ has the kinetic energy $W = \frac{1}{2} \cdot m_{veh} \cdot v_{coll}^{2}$. Consequently, the collision velocity criticality grows quadratically with the vehicle collision velocity. The criticality will have its maximum value $crit_{vel} = 1$ if the collision velocity is equal to the maximum domain speed, which was chosen to be $v_{max, domain}=\SI{30}{\kilo\metre\per\hour}$. For velocities below this border, the criticality can be determined by

	      \begin{align}
		      crit_{vel} & = \frac{v_{coll}^2}{v_{max, domain}^2}.
	      \end{align}

	\item Tube Criticality $crit_{tube}$: For each radar point, the distance from the tube is determined. The point receives a maximum score $c_{tube} = 1$, if it is located inside the driving tube. This is the case if the distance $d_{tube}$ to the center of the trajectory is smaller than half of the vehicle's width plus an additional safety zone of $\SI{0.1}{\metre}$. The safety zone is required to compensate insecurities from exact determination of the point's position as well as the error occurring during the path control. Outside of the tube, but close to it, the tube criticality function decreases in a so-called insecurity zone with a width of $\SI{2.0}{\metre}$ in the shape of a third-degree polynomial. This method guaranteed that the criticality score descends slowly and smoothly to zero.

	\item Distance Criticality $crit_{ttc}$: The higher the collision velocity $v_{coll}$ with a \gls{vru} is, the higher is $c_{dist}$. For the distance criticality $c_{dist}$, only the velocity component which the automated vehicle would contribute to the collision is considered - assuming that a deceleration action is immediately released. The required distance to stop is compound of a reaction time $t_{react}$ resulting in a reaction distance $d_{react} = v \cdot t_{react}$ and a stopping distance $d_{stop} = \frac{v^2}{2a}$ with the deceleration ability $a$.

\end{itemize}

These three criticality components are calculated for each radar point with its items $(x, y, v_{rel})$ in the unfiltered point cloud. Finally, the total point criticality $crit_{p}$ is calculated as a product of the three criticality components: velocity criticality $crit_{vel}$, tube criticality $crit_{tube}$ and distance criticality $crit_{ttc}$.

\begin{align}
	crit_{p} & = crit_{vel} \cdot crit_{tube} \cdot crit_{ttc}
\end{align}

We chose a simple multiplication since its identity element is $1$, which means that the total criticality is bounded to a maximum of $1$, and its zero property, implicating that absence of one criticality leads to a point which is not considered critical anymore ($crit_{p} = 0$). This fits to that a point is not critical if it is either far away, not in the drive tube or reached with a target velocity of \SI{0.0}{\metre\per\second}.

To derive actions from point criticality, two different ways result: the first one is to react continuously and adapt the intensity of the point processing to the criticality of the point. The second one is to define a criticality threshold $crit_{thresh}$, which divides radar points in the two categories "critical" and "not critical". A radar point $p$ exceeding this threshold $crit_{p} \geq crit_{thresh}$ is defined as a critical point. In this work, the second method is applied, which gives the advantage to derive a performance metric of the criticality evaluation by comparing true critical points with points evaluated as critical by the algorithm.

\subsection{Clustering}
\label{sec:cluster}

Point clouds are processed to objects using clustering and tracking. The clustering method groups multiple radar points which are in close relation to each other regarding their measurement vector, in particular the position and motion. The tracking algorithm merges the cluster centroids with the current object hypotheses by a nearest-neighbor approximation to update or generate motion state estimations. A created track can be considered an object hypothesis. However, tracks are created at the location of cluster centroids.

In this work, we consider a clustered radar point as the first, indispensable step to an object hypothesis and classify clustered points as correctly treated points. For this reason, tracking is no further part of the investigations of this work. The clustering is performed in a conventional way by a \gls{dbscan} approach without the application of machine learning methods or similar. This makes its behavior explainable. Details about the method are outlined in section \ref{sec:related_work}. The cluster selection distance for the nearest neighbor search is set to $\epsilon = 1.0$. With this parameter, the maximum distance of two points to count as member of one cluster can be adjusted. Choosing $\epsilon$ too low will result in a high number of clusters, even with more than one cluster on the same object. This would be unfavorable for the downstream tracking algorithm as each cluster evokes a track and tracking multiple times per cohesive object is unreasonable. However, a high cluster selection distance $\epsilon$ will lead to a conflation of two or more objects since a high distance is tolerated for two points still being members of the same cluster. Setting $\epsilon = 1.0$ results from several experiments with clustering which proved this value to be a sweet spot for \gls{vru} detection.

\subsection{Filtering}
\label{sec:filtering}

Common perception systems in vehicles feature multiple radar sensors, which are merged to receive one comprehensive point cloud for all subsequent operations. The purposes of this point cloud can be different, e.g., it is applied for motion estimation, localization, or object detection. To meet the prerequisites for these purposes, the number of points might be reduced in regions which are not considered relevant. As the simplest step, filtering cascades for the position $(x, y)$ may be contained. We assume for the object detection task that the surrounding objects move at a maximum of the domain speed $v_{domain, max} = \SI{30}{\kilo\metre\per\hour}$. The vehicle can easily be stopped in a distance of $x = \SI{20}{\metre}$ and lateral objects must only be observed up to $y = \SI{10}{\metre}$.

A further strategy especially in object detection is to filter for radar points which have been in a static motion condition. The velocity resolution of most automotive radars is around \SI{0.25}{\metre\per\second} \cite{Major2019}. A safety margin needs to be reserved for the odometry measurement, to compensate the ego motion of the measuring vehicle. Hence, in this work, we applied the condition $|v_{comp}| \geq \SI{0.5}{\metre\per\second}$. Here, the compensated velocity $v_{comp}$ is the measured doppler velocity of the point compensated by the velocity of the vehicle. Additionally, false radar points caused by, e.g., interference effects of other radar systems may occur in the system but can often be identified by unrealistic doppler velocity data. Especially in parking applications, no objects with an extremely high speed are expected in the driving domain. For this reason, the velocity filter may be set to $v_{dopp} \leq \SI{20.0}{\metre\per\second}$.

The points in the point cloud are now limited to a spatial rectangle and to a maximum doppler velocity, but the point cloud is still noisy. Thus, the downstream clustering algorithm might have problems to isolate objects from the noise floor. A common strategy to keep the most important radar points is to apply a filtering algorithm, which targets mainly the signal power or slightly converted the determined \gls{rcs} value per point. If the \gls{rcs} value of a radar point measure is low, this means that a small amount of energy is reflected. As the existence probability of these points is lower in general, the \gls{rcs} filter is a suitable method to avoid False Positives from noisy points. Matsunami et al. quantify a pedestrian's \gls{rcs} value as $\SI{-6}{\decibel\metre\squared}$ for a frequency of $\SI{77}{\giga\hertz}$ in worst case \cite{Matsunami2012}, while other works also determined values down to $\SI{-17}{\decibel\metre\squared}$ \cite{Chen2014}. Hence, in this work, we will investigate the false positive and false negative rate for different filter thresholds.

\subsection{Reachability Analysis}
\label{sec:reachability}

Facing actual critical scenes where a \gls{vru} is acutely in danger is a very seldom phenomenon. In most data sets for automated driving development, such scenes will rarely or not be found. Same applies to real driving data campaigns. An important objective of research is to obtain a large number of such situations without excessive data recording campaigns. Only sufficiently large situational sets can support to guarantee \gls{fit} rates of $\SI{1e-7}{\hour}$ or less. The current motion state and the intention of the automated driving system play a major role for the criticality of the driving scene. For instance, criticality is quickly deflected by braking or steering to a different direction. On the other hand, this means that choosing the trajectory purposefully will induce criticality.

As critical trajectories can barely be found in unmodified data sets, we propose to compute critical trajectories towards \gls{vru}s which serve as ground truth scenarios for the following experiments. These trajectories keep up the current vehicle dynamics, especially its current velocity $v_{0}$ and its yaw rate $\omega$. This means that they ensure a seamless transition between the real driving state and the fictional critical trajectory which a planner of an automated function could return as a planning result. The shape of the critical trajectory is always chosen to be a circular arc with a constant radius $r$ for simplicity. The notional collision speed $v_{coll}$ can be chosen arbitrarily with respect to the feasible lateral and longitudinal vehicle dynamics.

\begin{align}
	r        & = \frac{x_{vru}^{2} + y_{vru}^{2}}{2y_{vru}}                    \\
	t_{coll} & = \frac{2r\arcsin{\frac{x_{vru}}{r}}}{v_{0} + v_{coll}} - t_{0}
\end{align}

In the experiments, the maximum longitudinal acceleration which is still drivable is $a_{long, max} = \SI{10.0}{\metre\per\second\squared}$ while the lateral acceleration was considered to be limited at $a_{lat, max} = \SI{8.0}{\metre\per\second\squared}$. Besides, trajectories whose radius falls below the minimum feasible radius of $r = \SI{6.0}{\metre}$ are not recognized as reachable trajectories.

\begin{figure}
	\begin{subfigure}[b]{.24\textwidth}
		\centering
		\includegraphics[width=\linewidth]{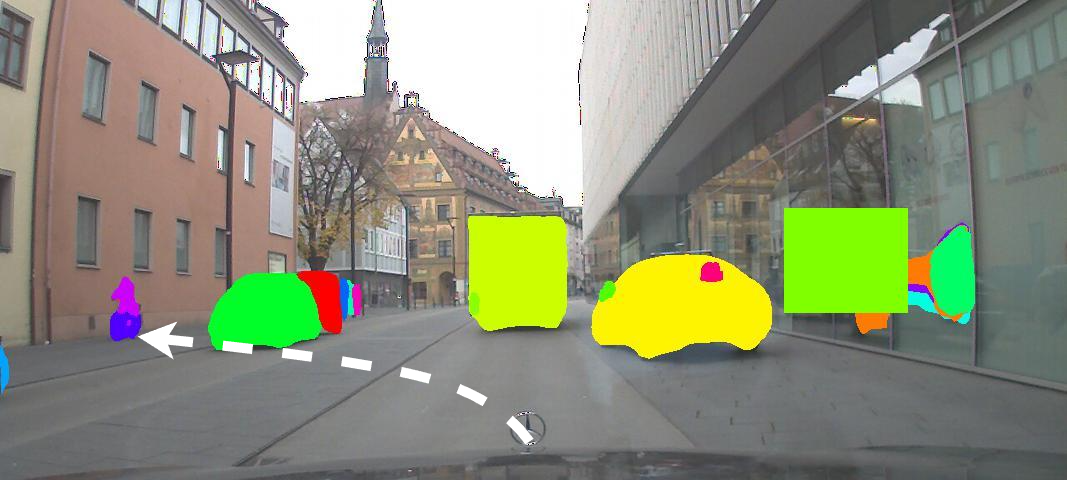}
	\end{subfigure}
	\hfill
	\begin{subfigure}[b]{.24\textwidth}
		\centering
		\includegraphics[width=\linewidth]{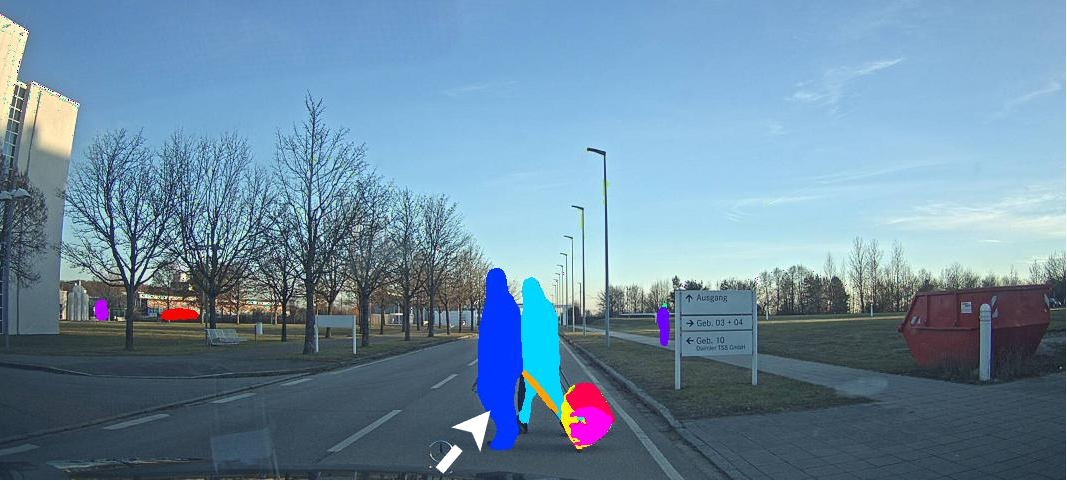}
	\end{subfigure}

	\medskip

	\begin{subfigure}[t]{.24\textwidth}
		\centering
		\includegraphics[width=\linewidth]{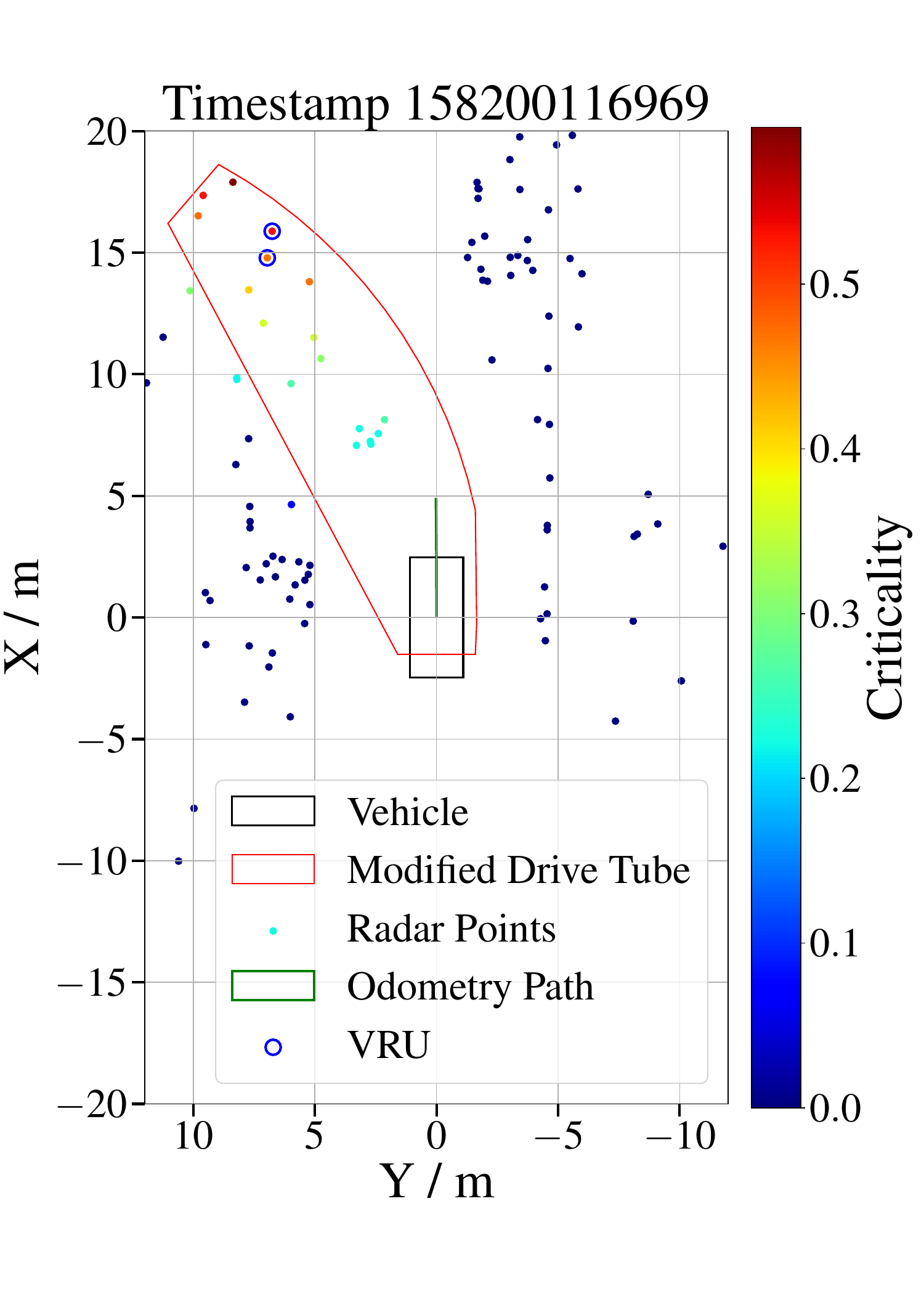}
	\end{subfigure}
	\hfill
	\begin{subfigure}[t]{.24\textwidth}
		\centering
		\includegraphics[width=\linewidth]{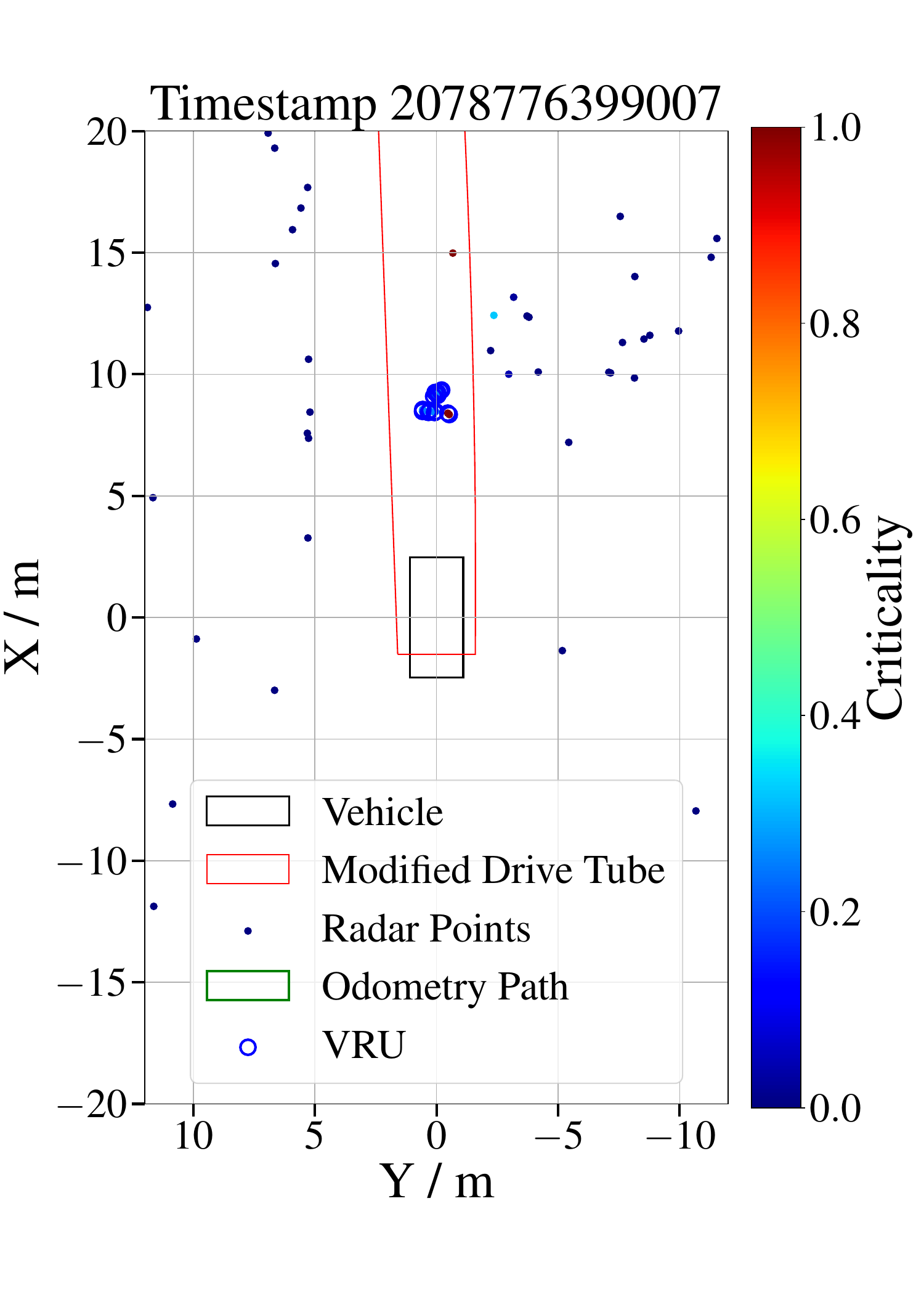}
	\end{subfigure}
	\caption{Two types of scenarios to create a hazardous situation with \gls{vru}s}
	\label{fig:reachability}
\end{figure}

In usual, non-critical driving scenarios, two classes of scenarios to encounter \gls{vru}s can be named, which are depicted in figure \ref{fig:reachability}. The color of the radar points shows our criticality evaluation. The points which represent the \gls{vru} are highlighted by a blue circle around.

In the first scenario class, \gls{vru}s move next to or on the pedestrian walk. An example of this class is shown on the left-hand side of the figure. To reach the bicyclist on the opposite lane, the yaw rate $\varphi$ of the vehicle must be modified to let the vehicle turn to the left. It differs from the green path which is the actual motion path of the vehicle in this scene.

In the second scenario class, \gls{vru}s are actively crossing or moving in front of the vehicle. One scenario of this class is depicted on the right-hand side of the figure. The vehicle is actually in stand still, so an acceleration $a$ must be applied to induce criticality on these \gls{vru}s. The modified drive tube is straight, but the motion state is constantly accelerating.

As the investigation takes part targeting the parking context, the maximum speed of the trajectory is set to $v_{max, domain} = \SI{8.5}{\metre\per\second}$. While the adapted trajectory increases the velocity to produce a scenario which is as severe as possible, the hypothetical acceleration at $v_{max, domain}$ is set to zero, meaning that the trajectory aims at a constant drive.

Note that to find radar points representing \gls{vru}s which suit as a target for a critical trajectory on a virtual collision route, a data set with class labels per radar point is required.

\subsection{Point Treatment}
\label{sec:treatment}

\begin{figure}[ht]
	\centering
	\includegraphics[width = \linewidth]{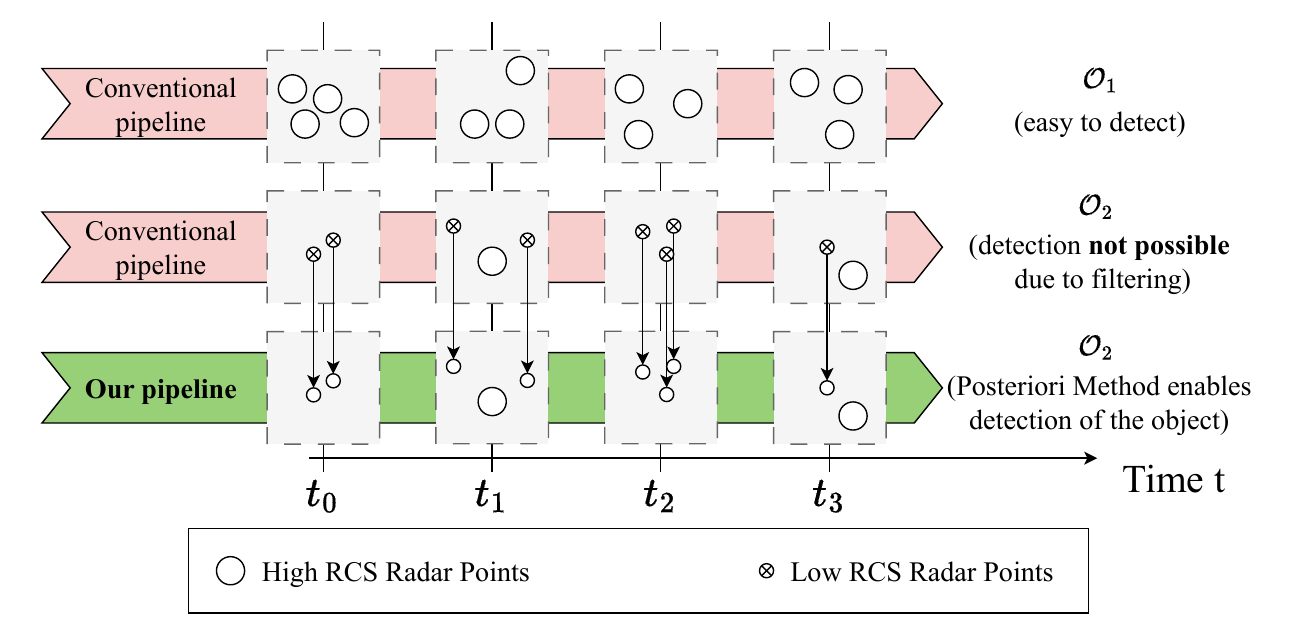}
	\caption{Comparison of radar points on object $\mathcal{O}_{1}$ which is easy to detect, object $\mathcal{O}_{2}$ which is difficult to detect, and our approach which improves availability of information about weakly reflecting objects}
	\label{fig:detection_pts}
\end{figure}

In the previous subsections, we highlighted the fact that truly critical points might be deleted by conventional filters. Figure \ref{fig:detection_pts} shows the difference between an object $\mathcal{O}_{1}$ which can be detected easily and one object $\mathcal{O}_{2}$ which is tough to detect. While $\mathcal{O}_{1}$ is represented by enough points with a high \gls{rcs},  value, the object $\mathcal{O}_{1}$ features less detections with a low \gls{rcs} value and is not detected at all in some cycles. A conventional filtering algorithm would remove the detections with the low \gls{rcs} value so that all remaining information about the object is completely lost. Our approach sharpens the prospective awareness in regions with critical radar points. It lowers the filter threshold in these regions and achieves that points with a low \gls{rcs} value still stay available for the downstream algorithms.

\begin{figure}[ht]
	\centering
	\includegraphics[width = \linewidth]{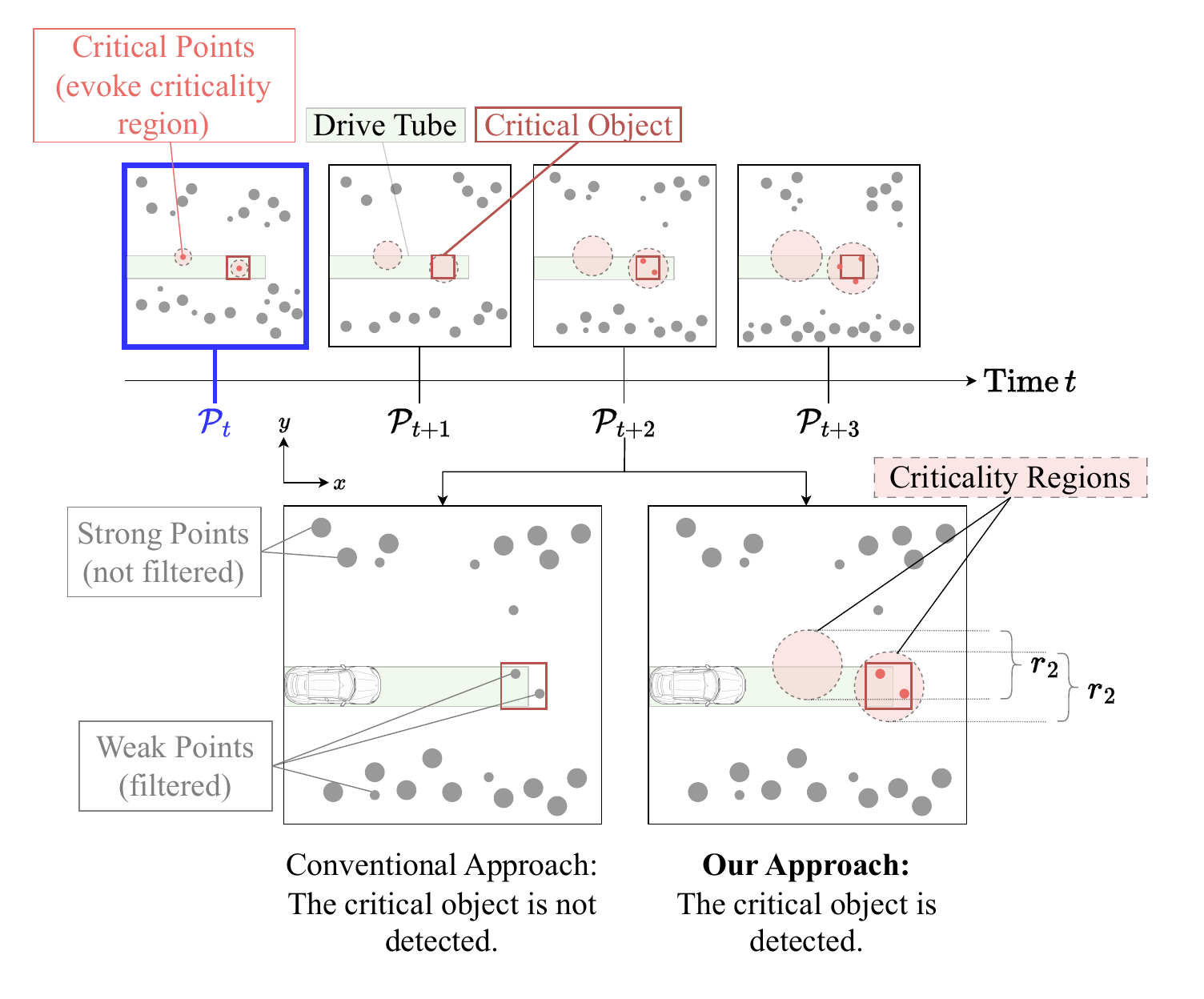}
	\caption{Posterior method to adapt point processing in a safety-aware manner}
	\label{fig:posterior_pt_treatment}
\end{figure}

Other ways which could be applied in critical regions are the recovery of information which was originally removed or adding more information achieved by other sensors to the critical points. Due to the confined scope, this work did not focus on these approaches. The concept of lowering the future threshold (\textit{posteriori}) specifically in certain regions is shown in figure \ref{fig:posterior_pt_treatment}.

By analyzing these critical regions in consecutive point clouds, it can be guaranteed that a point in regions with evidence on object existence will not be deleted wrongly. We call the method \textit{Posteriori Method}, as the following point clouds with the horizon length $N_{posterior}$ are processed with a reduced threshold in critical regions. These criticality regions are defined for the following measurements $\mathcal{P}_{t+1}, \dots, \mathcal{P}_{t+N_{posterior}}$ around each critical point of the current point cloud $\mathcal{P}_{t}$. In figure \ref{fig:posterior_pt_treatment}, the criticality regions are the reddish circles which evoke around the critical points which are depicted as orange dots. In the criticality region, the filter criteria are adapted. We propose to completely remove the filter criterion for the \gls{rcs} value of each radar point which lies inside of a criticality region. This implicates that all points in the criticality region will be included during the next point clouds, independently of their \gls{rcs}, as can be seen in figure \ref{fig:posterior_pt_treatment} for the point cloud $\mathcal{P}_{t+2}$ exemplarily.

The regions are not tracked as the motion state and direction of a single point cannot be determined if no tracking algorithms are applied. However, the regions grow as the position which was originally measured is subject to measurement noise processes and the object represented by the point may be in motion. Additionally, the regions are transformed to compensate the ego motion of the measuring vehicle. The regions have a lifetime $t_{life}$ which is set to $t_{life} = 5$ cycles to keep the number of regions reasonably low and ensure that the time looking for object existence evidence is limited. The regions are parameterized by the radius $r$. We let this radius grow with higher region age.

%For the first concept mentioned, it is required to restore information which has originally not been investigated during the signal processing. More precisely, a horizon with a length of $N_{priori}$ measurements is introduced. All these point clouds are stored in a circular buffer. If a critical point occurs in the measurement $m_{t}$, the preceding point clouds $\mathcal{P}_{t-1}, \dots, \mathcal{P}_{t-N_{priori}}$ are investigated retrospectively in the regions of the critical point occurrence. These so-called awareness regions around critical points stay static in a world-fixed coordinate system. This means that a transformation is required in a sensor or vehicle coordinate system as soon as the vehicle is moving. The filter criterion can either be undermined or even suspended in identified regions. This leads to point clouds $\tilde{\mathcal{P}_{t-1}}, \dots, \tilde{\mathcal{P}_{t-N_{priori}}}$ which contain not only the points of the filtered point cloud, but also points which have been deleted formerly due to there low signal power or \gls{rcs}. The resulting point clouds are then processed by the clustering and tracking algorithm. As a result, an extended set of tracks is received which takes account of critical regions which have evidence on object presence. 

\section{EXPERIMENTAL SETUP}
\label{sec:experimental_setup}

The purpose of the experiments is to evaluate the performance of the proposed criticality evaluation function as well as to prove that the critical points which are identified can be treated in such a way that the deletion of radar points representing objects is prevented as much as possible without a high increase of false positives, i.e., object hypotheses formed by noise points. By that, we aim to show that the pipeline introduced in this work can increase the safety of an automated function significantly in comparison with recent SAE level 2 perception mechanisms. For this, a state-of-the-art perception algorithm is used as a reference. This algorithm includes a conventional filter structure as described in section \ref{sec:filtering}, which deletes radar points with a low \gls{rcs} value. These points reflected low signal power and are hence anticipated to feature a low existence probability. This principle is used similarly in many SAE level 2 functions. Moreover, clustering based on \gls{dbscan} is performed to group the points. This algorithm proved to work well in almost all cases, although it accepts to miss objects from time to time for the sake of keeping brake actions induced by false positives rare. In other words, the recall is lowered to increase the precision metric of the perception algorithm, which enhances the overall robustness of the function.

In the following, three quantities $A$, $B$, and $C$ are defined. The quantity $A$ comprises all points which are classified as critical by our driving tube-based criticality evaluation method. The quantity $B$ comprises all points which are not identified by the reference algorithm and can hence lead to critical situations if they are not treated. The set $B$ can be considered as the ground truth set of all truly critical radar detections. In this work, the quantity $B$ is received artificially by applying the reachability method of section \ref{sec:reachability}. The quantity $C$ comprises all points which were structured in a cluster by the \gls{dbscan} clustering algorithm, which means that eventually an object was formed from the clustering centroid.

Ideally, the quantity $A$ comprises all points of the quantity $B$. In this case, all relevant, true-critical points are identified by the proposed criticality evaluation algorithm. The points $A \cap \bar{B}$ are points with a high criticality score which have indeed a low criticality. The objective is to reduce these points, e.g., to reduce the frequency of false brake actions. The points $\bar{A} \cap B$ show where the proposed algorithm still fails and must be carefully investigated. These points are true critical points which are not identified as critical by the criticality evaluation method.

\begin{figure}[ht]
	\centering
	\includegraphics[width = \linewidth]{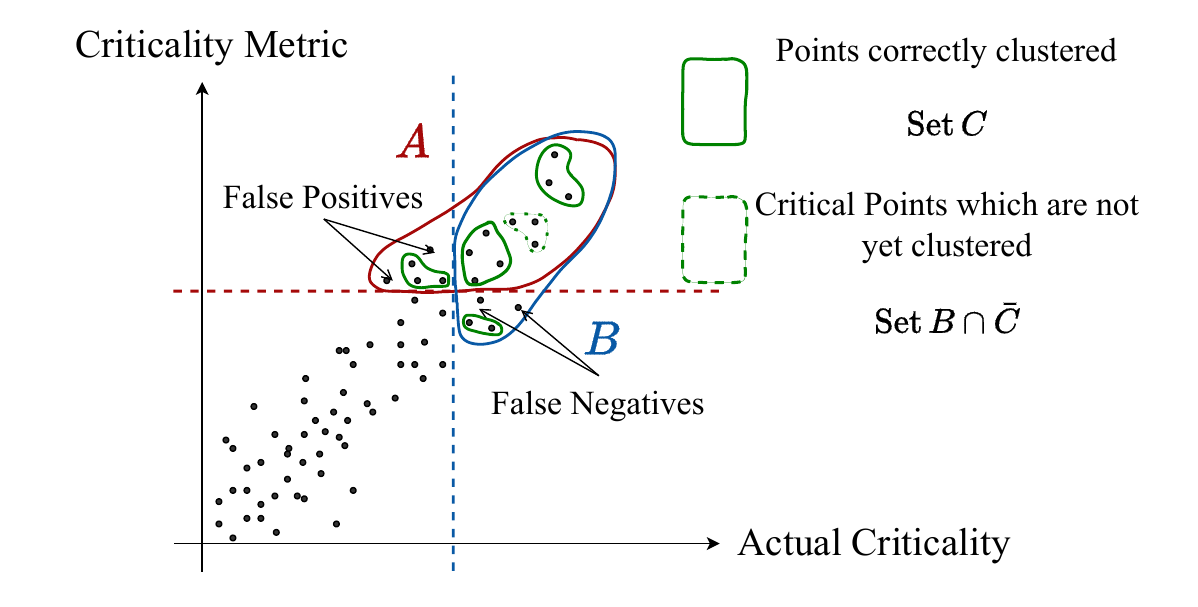}
	\caption{Sets $A$, $B$, and $C$ to determine the reduction of critical situations}
	\label{fig:sets_a_and_b}
\end{figure}

Figure \ref{fig:sets_a_and_b} gives an overview about the quantities. It can be seen that points exist which hold $\bar{A} \cap B \cap C$. In this case, although the critical points were not identified as such, the processing algorithm formed a cluster. These points are relevant for the improvement of the criticality evaluation method, but not for the evaluation of critical point post-processing. Hence, only points in set $B$ which have not been clustered are considered \textit{false negatives}.

For the first experiment, the evaluation performance of criticality is investigated by using the RadarScenes data set \cite{Schumann2021}. Its data is split in 158 sequences with each featuring a length of \SI{16.71}{\second} to \SI{285.40}{\second}. The total distance of the recordings is \SI{92.00}{\kilo\metre} which was driven in a time of \SI{4.45}{\hour}.

This data set was chosen because it features dense radar point clouds with enough points on \gls{vru}s as well as noise points. Another advantage is that RadarScenes point clouds were labeled semantically per radar point. As the scope of the investigations lies on automated parking functions, the main hazard during operation is if the automated vehicle collides with these \gls{vru}s. Whereas a collision with traffic participants such as cars or trucks would primarily lead to a loss of property, which can be prioritized lower in safety-aware considerations. For this reason, only scenes with low speed domains were analyzed in the experiment, where the driven velocity of the automated vehicle does not exceed $v=\SI{30} {\kilo\metre\per\hour}$.

The parameters of the \textit{posteriori} method are subject to the measurement frequency, resp., the measurement period. Specifically, the region sizes for every region age step needs to be selected. For the experiment, the region radius sizes $r_{1} = \SI{0.2}{\metre}$, $r_{2} = \SI{0.4}{\metre}$, $r_{3} = \SI{0.6}{\metre}$, $r_{4} = \SI{0.8}{\metre}$, and $r_{5} = \SI{1.0}{\metre}$ were set. With a cycle period of approximately $\SI{0.091}{\second}$, this growth covers absolute object speeds of up to $v_{obj} = \frac{\SI{0.2}{\metre}}{\SI{0.091}{\second}} = \SI{2.2}{\metre\per\second}$, which is a pedestrian walking quickly. Higher growth rates can be selected, but covering faster \gls{vru}s will be at the cost of robustness and increasing number of false positives.

At first, the data set was evaluated to analyze the number of critical scenarios, using the current motion state of the vehicle propagated for the next seconds as the driving tube for evaluation. It was found that there are hardly any critical radar scenes in the data set. This is not a weak point of RadarScenes, but most of the data sets for the development of automated driving functions do not feature critical scenes. Since criticality emerges from the driving tube, an easy way to create artificially critical scenes is the manipulation of the planned trajectory. The critical trajectory is found using the method outlined in section \ref{sec:reachability}. Once a reachable point $p_{\textit{reach}}$ is found which belongs to one of the semantic classes [bicycles, pedestrians, group of pedestrians], this point is added to set $B$ and the search is terminated. However, all points with the same \textit{track id} as the point $p_{\textit{reach}}$ are also classified as critical points and added to set $B$, as far as the Euclidean distance to the point $p_{\textit{reach}}$ is less than \SI{2.0}{\metre}. This prevents that numerous points of a huge entity are in set $B$, although the originally determined trajectory is not able to reach all points at once.

\begin{figure}
	\centering
	\begin{subfigure}[b]{0.35\textwidth}
		\centering
		\includegraphics[width=\linewidth]{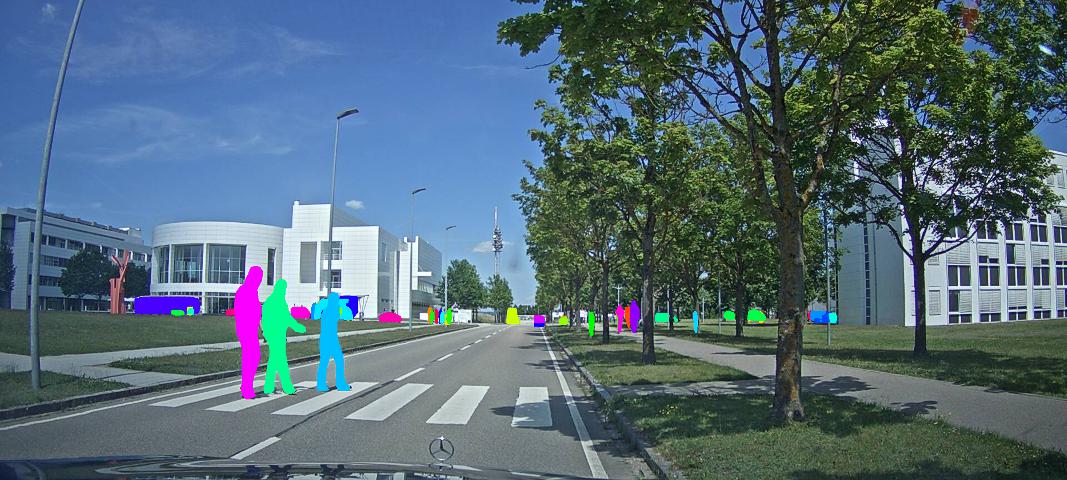}
	\end{subfigure}
	\medskip
	\vspace{0.3cm}
	\begin{subfigure}[b]{.35\textwidth}
		\centering
		\includegraphics[width=\linewidth]{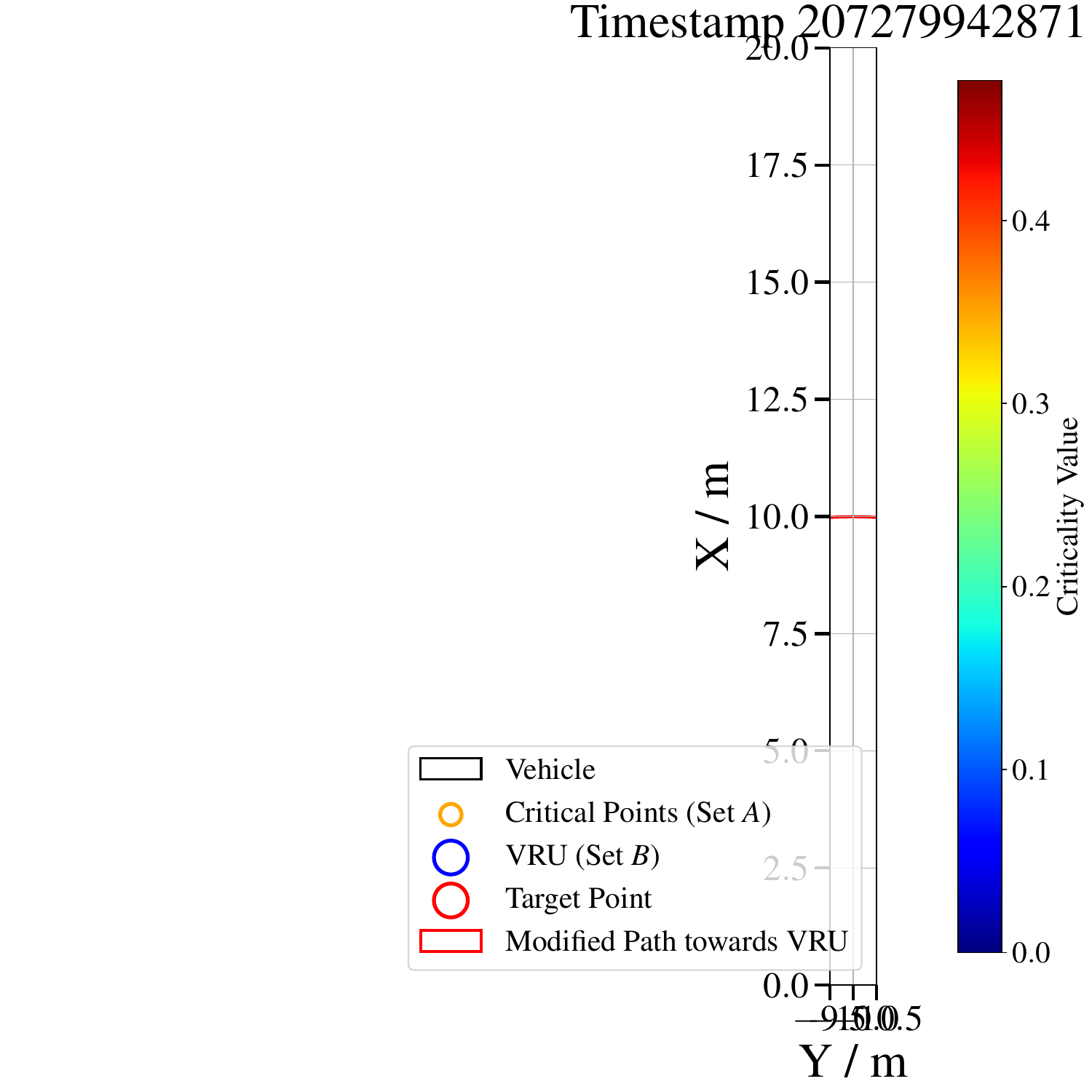}
	\end{subfigure}
	\caption{Scene in sequence 63 with modified path towards pedestrians. Orange points are evaluated critical (Set $A$). Points with a blue circle around are pedestrians, as can be seen in the camera image and ground truth critical points (Set $B$)}
	\label{fig:artif_critic}
\end{figure}

Figure \ref{fig:artif_critic} shows the original situation and how the number of critical points increased by changing the driving tube.

% For the real world approach, expert knowledge was used to create a clean, artificial scene. One main question during the experimental design was how ground truth criticality can be defined. During the data set experiment, critical objects were selected manually by experts, using the information from the documentation camera as well as the radar point labels. Now, two different approaches could be taken. The first one would be to record the driver’s view and defining those points as critical which could be resolved by the driver and are not handled correctly by the reference algorithm. The second approach would incorporate more sensors, such as additional cameras, lidar sensors, \gls{rtk} or an infrastructure camera. In case of the first method, this means that the reflections from objects which are not seen by a human driver would not count as fail objects if they are not handled by the conventional algorithm. In the second case, the demands on the perception function are significantly higher, since it needs to handle each arising function which is critical and identified by the omniscient sensor-infrastructure construct needs to be handled by the pipeline as well.

%\begin{itemize}
%\item Reactive Clustering and Tracking: Add preliminarily filtered, but critical points to the processing pipeline
%\item Critical Point Einrichment: Add critical points directly as a clustering center
%\end{itemize}

\section{EXPERIMENTAL RESULTS}
\label{sec:experimental_results}

% Hint: the evaluation of this paper was made with the scripts in Jenny's branch of the scenarios pkg.
% -"analyse_set_A.py" for the recall, precision, f1 score and the figure of the threshold vs. the metric scores.
\subsection{Results of the Criticality Evaluation}
\label{sec:crit_eval_results}

A first step in the experimental evaluation is the analysis of the individual sequences. Critical scenes by definition are scenes in which a \gls{vru} is compromised by the vehicle in a way that a collision with injury occurs if no action is undertaken. As standard, the condition is determined by the current vehicle state, which is essentially defined by the current velocity $v_{x}$ of the vehicle and its yaw rate $\omega$. Manual inspection of the data set showed that this situation does not occur in the data set. Consequently, to fill set $B$ with points, the trajectory is adapted. The premise for usage of the sequence is that at least one \gls{vru} emerges in the sequence and the trajectory can be adapted to reach this \gls{vru} and induce a critical scene. In total, 109 of the 158 sequences could be used as there was at least one reachable critical point of set $B$, which results in a percentage of \SI{69.0}{\percent}. The total number of points in set $B$ after trajectory adaption is \num{187106} which are distributed over \num{24885} point clouds. The trajectory is modified during \SI{12.6}{\percent} of all timestamps.

For the analysis, the points which did not end in a cluster are of high interest. In the following, we use the symbol $F$ to denote the set of all points that are members of the point cloud filtered by the conventional \gls{rcs}-based algorithm. These points might miss the cluster due to two different reasons: either they have been deleted by the conventional filtering algorithm ($B \cap \bar{F}$), or they weren't clustered since the number of neighboring points was not sufficient to form a new cluster ($B \cap F \cap \bar{C}$). This in turn can happen due to other weak points which were filtered and reduce the point density vitally. Another reason is a low point density on the object. In the latter case, increasing the parameter $\epsilon$ can help to still form a cluster. On the same hand, we could omit the cluster requirement for an object hypothesis. Instead, it could be based on a single point if this point is critical enough. Note that these considerations should be handled with care since they could cause a high number of false positive object hypotheses and they are not further discussed in this work.

To be considered in the post processing step, the radar point needs to exceed the criticality threshold $\textit{crit}_{\textit{thresh}}$ which adds it to set $A$. To find the threshold, a reasonable balance shall be found between including all critical points and keeping a false positive clustering rate low. Precision and recall have been evaluated and additionally, the F\textsubscript{1} score as the harmonic mean between both indicators is determined.

\begin{align}
	\textit{Recall}      & = \frac{A \cap B}{B}                                                                            \\
	\textit{Precision}   & = \frac{A \cap B}{A}                                                                            \\
	F_{1} \textit{score} & = \frac{2 \cdot \textit{Recall} \cdot \textit{Precision}}{\textit{Recall} + \textit{Precision}}
\end{align}

\begin{figure}[ht]
	\centering
	\includegraphics[width = \linewidth]{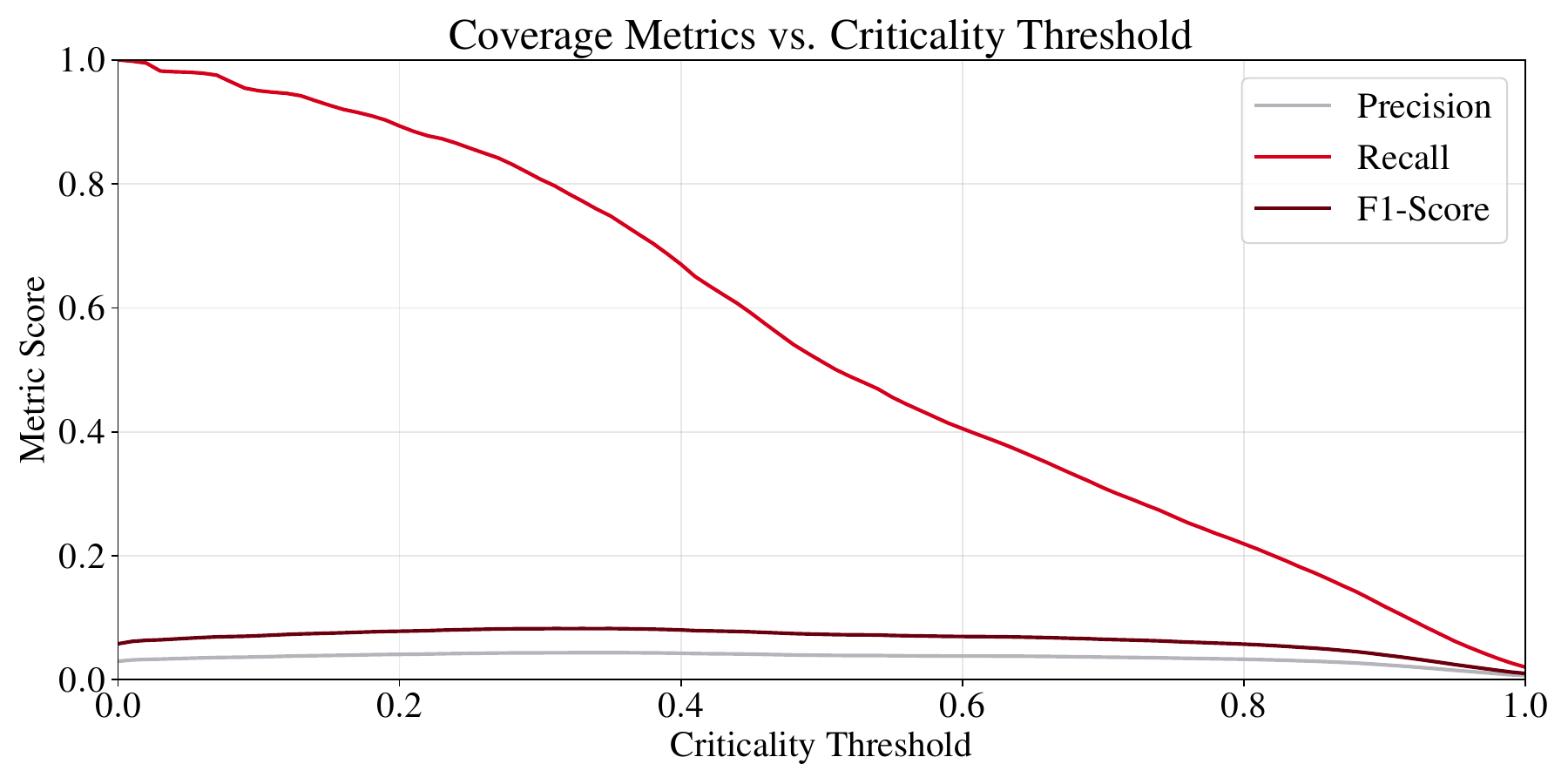}
	\caption{Recall, Precision, and F\textsubscript{1} score for different threshold values $\textit{crit}_{\textit{thresh}}$}
	\label{fig:recall_precision_thresh}
\end{figure}

The distribution of the criticality evaluation over the whole data set is shown in figure \ref{fig:recall_precision_thresh}. The recall line is decreasing slightly until a threshold of $0.3$ and then decreases more steeply. This means that true critical points of set $B$ more often receive a criticality score of $crit_{p} > 0.3$, however, a notable amount receives only a minor criticality value. The reason behind this is that some points of set $B$ are close to the vehicle and the collision velocity $v_{coll}$ with these points is comparably low. The data set structure of RadarScenes does not allow to find a ground truth value for the severity of injuries in case of a fictitious collision.

As a primary objective was to increase the safety, a high recall value ranks first. Hence, we decided to perform the experiment with two threshold values, one which is a safety-aware setup ($\textit{crit}_{\textit{thresh}} = 0.1$) and one at the optimum F\textsubscript{1} score ($\textit{crit}_{\textit{thresh}} = 0.35$), which balances best between recall and precision. The key data of both thresholds is listen in table \ref{tab:safe_vs_f1opt}. Note that independently of the criticality threshold parameter, it is impossible to achieve a significant precision. The reason for this is that the criticality evaluation metric is not able to distinguish between \gls{vru}s and reflections from other objects. In general, only a minor part of \SI{1.20}{\percent} of the total number of radar points belong to one of the \gls{vru} classes. From this point of view, a low precision seems reasonable for a criticality evaluation without object classification.

\begin{table}[ht]
	\caption{Comparison of two different criticality threshold values $\textit{crit}_{\textit{thresh}}$}
	\centering
	\label{tab:safe_vs_f1opt}
	\begin{tabular}{cccc}
		\toprule
		                         & $\textit{crit}_{\textit{thresh}} = 0.1$ & $\textit{crit}_{\textit{thresh}} = 0.35$ \tabularnewline
		\midrule
		$\sum$ Points in set A   & \num{4086920}                           & \num{2716981} \tabularnewline
		\midrule
		Recall                   & \SI{95.02}{\percent}                    & \SI{74.81}{\percent} \tabularnewline
		\midrule
		Precision                & \SI{3.69}{\percent}                     & \SI{4.37}{\percent} \tabularnewline
		\midrule
		F\textsubscript{1} score & \SI{7.10}{\percent}                     & \SI{8.26}{\percent} \tabularnewline
		\bottomrule
	\end{tabular}
\end{table}

It can be recognized that the recall at $\textit{crit}_{\textit{thresh}} = 0.1$ is at a high level and the F\textsubscript{1} score is only $1.16$ percentage points below the optimum. As a safety-aware setup was important for our field of application, we decided to continue our evaluation with this setting as a threshold.

\subsection{Results of the Baseline Clustering}
\label{sec:bl_cluster_results}

We performed experiments with four different filter thresholds. The strongest filter removes all points with a \gls{rcs} smaller than \SI{0.0}{\decibel\metre\squared}, while the weakest one removes only very weak reflections with a \gls{rcs} of smaller than \SI{-15.0}{\decibel\metre\squared}. The filters and the share of radar points being filtered in the set can be seen in table \ref{tab:rcs_filters}.

\begin{table}[ht]
	\caption{Different filter thresholds with their filter rate}
	\centering
	\label{tab:rcs_filters}
	\begin{tabular}{ccc}
		\toprule
		Threshold                          & $\sum p_{\textit{remaining}}$ & Filter Rate $\frac{\sum p_{\textit{remaining}}}{\sum p_{\textit{total}}}$ \tabularnewline
		\midrule
		\SI{0.0}{\decibel\metre\squared}   & \num{1448442}                 & \SI{1.44}{\percent}\tabularnewline
		\midrule
		\SI{-5.0}{\decibel\metre\squared}  & \num{2840936}                 & \SI{2.83}{\percent} \tabularnewline
		\midrule
		\SI{-10.0}{\decibel\metre\squared} & \num{5160073}                 & \SI{5.14}{\percent} \tabularnewline
		\midrule
		\SI{-15.0}{\decibel\metre\squared} & \num{9319639}                 & \SI{9.28}{\percent} \tabularnewline
		\bottomrule
	\end{tabular}
\end{table}

\begin{figure}[ht]
	\centering
	\includegraphics[width = \linewidth]{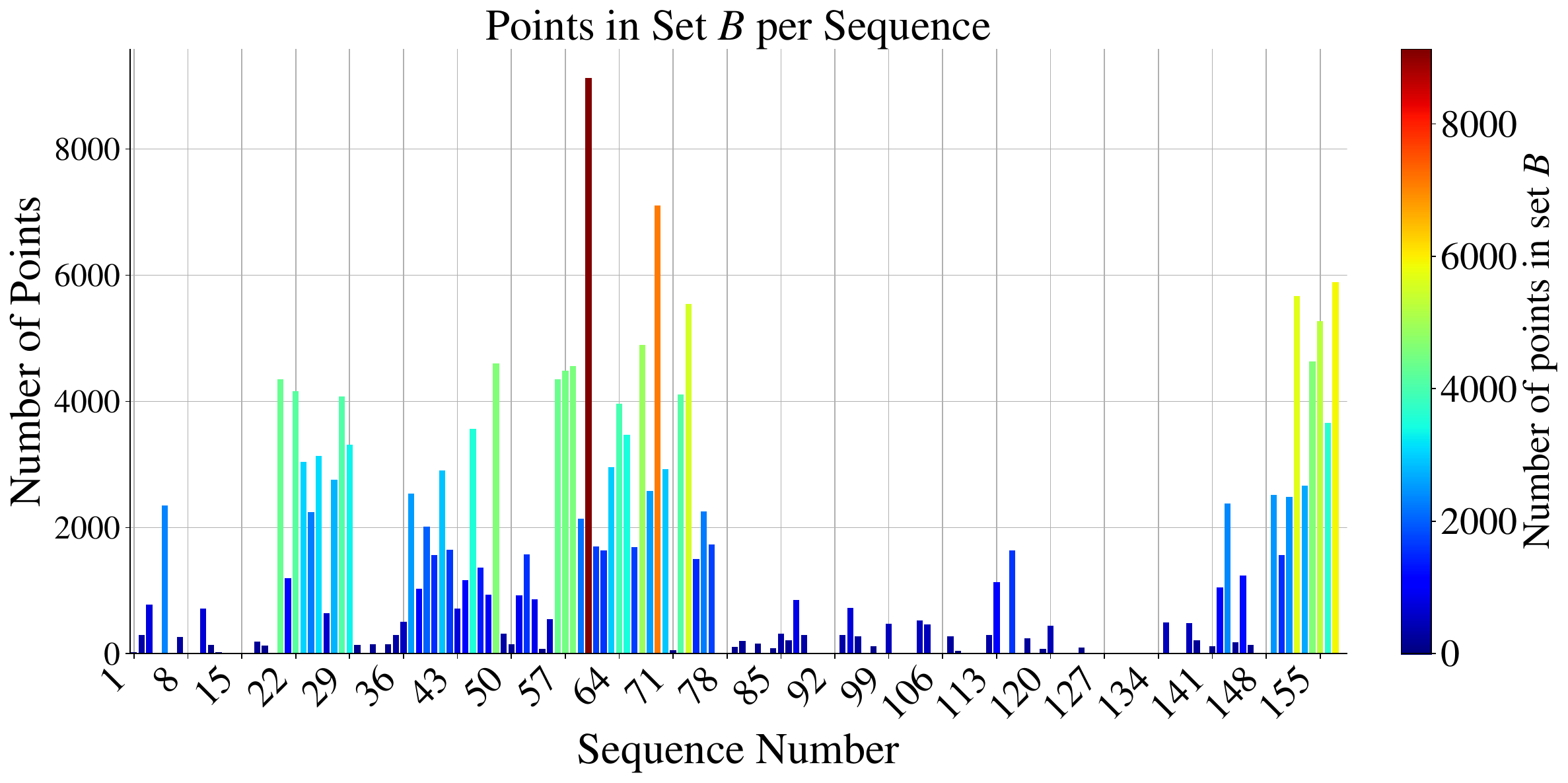}
	\caption{Distribution of critical radar points (set $B$) per RadarScenes sequence}
	\label{fig:b_points_per_sequence}
\end{figure}

The filter design was chosen to be relatively aggressive to keep only objects which are relevant for a parking planning task. Note that \gls{vrus} in a far distance do not appear in set $B$, as they do most likely not fulfill the reachability criterion. Figure \ref{fig:b_points_per_sequence} shows the distribution of set B points for each sequence. As can be seen, the number of points vary heavily between the sequences.

To determine the basic performance of the object detection algorithm, a few indicators will be compared. The most important one is the number of ground truth critical points, which have not been clustered ($B \cap \bar{C}$). In contrast to this number, the robustness can be determined by the number of points which have totally been clustered (set $C$). Ideally, no cluster should be created for noise points, while clustering existing static objects such as lanterns, pillars, guard rails etc., is a desirable behavior. Unfortunately, RadarScenes does not distinguish between static objects and noise points. However, a distinguishing label would be required to derive a meaningful metric for false positives. The only way to avoid this is to evaluate only moving objects. As the performance for standing \gls{vru}s is highly relevant, this method was discarded.

The results for the experiments without the safety-aware point treatment are listed in table \ref{tab:results_wo_treatment}. The percentage in column $B \cap C$ refer to the percentage of clustered points in all truly critical points of set $B$.

\begin{table}[ht]
	\caption{Clustering performance without safety-aware treatment}
	\centering
	\label{tab:results_wo_treatment}
	\begin{tabular}{ccccc}
		\toprule
		Threshold                          & $\sum C$      & $B \cap C$                         & $B \cap \bar{F}$ & $B \cap \bar{C}$ \tabularnewline
		\midrule
		\SI{0.0}{\decibel\metre\squared}   & \num{163018}  & \num{77} (\SI{0.04}{\percent})     & \num{185203}     & \num{1085} \tabularnewline
		\midrule
		\SI{-5.0}{\decibel\metre\squared}  & \num{595928}  & \num{783} (\SI{0.42}{\percent})    & \num{176752}     & \num{8930} \tabularnewline
		\midrule
		\SI{-10.0}{\decibel\metre\squared} & \num{1721085} & \num{13039} (\SI{7.00}{\percent})  & \num{146314}     & \num{27040} \tabularnewline
		\midrule
		\SI{-15.0}{\decibel\metre\squared} & \num{4281094} & \num{60165} (\SI{32.29}{\percent}) & \num{98530}      & \num{27651} \tabularnewline
		\bottomrule
	\end{tabular}
\end{table}

We observed that the aggressive filtering removes a huge part of all points - even in the mildest filter threshold setting (\SI{-15.0}{\decibel\metre\squared}), over \SI{55}{\percent} of all ground truth critical points have been removed by the filter (category $B \cap \bar{F}$). Each of these points can entail an undetected object, which is in danger if the planner does not take it into consideration. To guarantee safety, the major objective is to detect all objects at least in the drive tube. If no safety-aware approach such as the one proposed by us is chosen, the only chance to obtain more critical radar points is a very mild filter, which is prone to yield a high number of false positives.

\subsection{Results of the Posteriori Processing Method}
\label{sec:posteriori_results}

In the next step, the results of the experiments with our safety-aware filtering algorithm are shown. Table \ref{tab:pc_filter_comparison} compares key indicators of the detection performance without and with the modified pipeline, including the safety-aware processing.

\begin{table*}[ht]
	\caption{Clustering performance comparison without and with safety-aware processing}
	\centering
	\label{tab:pc_filter_comparison}
	\begin{tabular}{ccccccc}
		\toprule
		Threshold                          & \makecell{Filter Rate                                                                                                                                                                                              \\(without)} & \makecell{Filter Rate \\ (with)} & \makecell{Correctly treated \\$B \cap C$ (without)} & \makecell{Correctly treated \\$B \cap C$ (with)} & \makecell{Incorrectly filtered \\$B \cap \bar{F}$ (with)} & \makecell{Clustered points \\ $\sum C$ (with)} \tabularnewline
		\midrule
		\SI{0.0}{\decibel\metre\squared}   & \SI{1.44}{\percent}   & \SI{6.83}{\percent}  & \num{77} (\SI{0.04}{\percent})     & \num{142699} (\SI{76.30}{\percent}) & \num{31392} (-\SI{83.0}{\percent}) & \num{3961374} (+\SI{2330}{\percent})\tabularnewline
		\midrule
		\SI{-5.0}{\decibel\metre\squared}  & \SI{2.83}{\percent}   & \SI{8.01}{\percent}  & \num{783} (\SI{0.42}{\percent})    & \num{142862} (\SI{76.33}{\percent}) & \num{30929} (-\SI{82.5}{\percent}) & \num{4337161} (+\SI{628}{\percent})\tabularnewline
		\midrule
		\SI{-10.0}{\decibel\metre\squared} & \SI{5.14}{\percent}   & \SI{10.04}{\percent} & \num{13039} (\SI{7.00}{\percent})  & \num{143312} (\SI{76.60}{\percent}) & \num{30004} (-\SI{79.5}{\percent}) & \num{5339292} (+\SI{210}{\percent})\tabularnewline
		\midrule
		\SI{-15.0}{\decibel\metre\squared} & \SI{9.28}{\percent}   & \SI{13.66}{\percent} & \num{60165} (\SI{32.29}{\percent}) & \num{144934} (\SI{77.46}{\percent}) & \num{28383} (-\SI{71.2}{\percent}) & \num{7601955} (+\SI{77.6}{\percent})\tabularnewline
		\bottomrule
	\end{tabular}
\end{table*}

\begin{figure}[ht]
	\centering
	\includegraphics[width = \linewidth]{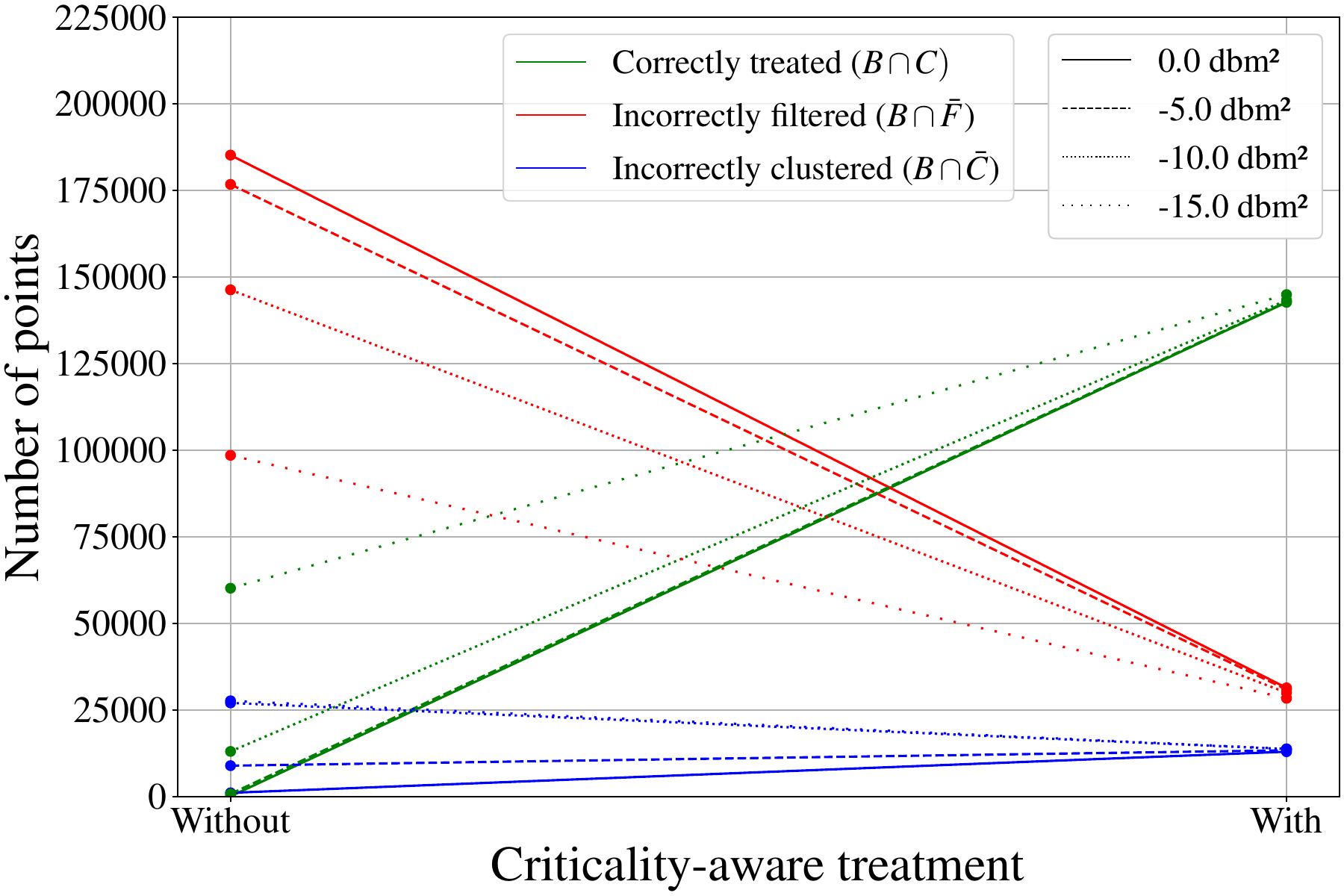}
	\caption{Comparison of false filtering, false clustering, and correct treatment without and with safety-critical treatment}
	\label{fig:compare_processing_fig}
\end{figure}

As weak points of criticality regions are added to the filtered point cloud, the number of points after the filtering increases significantly by the safety-aware processing. For the most rigorous threshold (\SI{0.0}{\decibel\metre\squared}), the number of processed points is nearly five times the number in the experiment without the safety-aware processing. Next, it stands out that the number of correctly treated points is increased almost independently of the previously selected filter threshold at a robust $76-\SI{77}{\percent}$. This proves that the initially selected threshold is of secondary importance and high filter rates are admissible without risking high false negative rates. The increase of clustered points is evidence for a denser point cloud which succeeds to map more points to existing clusters and generates new clusters. It can be assumed that most of the new clusters are generated in the driving tube of the vehicle. This implies that the probability for critical scenes will also decrease significantly, since false negative objects will only appear from incorrectly filtered radar points ($B \cap \bar{F}$). The improvement is shown in figure \ref{fig:compare_processing_fig} in more detail.

This figure complements the data shown in table \ref{tab:pc_filter_comparison}. The number of incorrectly filtered points is reduced, while the number of correctly clustered points increased significantly. Interestingly, the number of non-clustered points increases in some experiments. The reason for this is that some critical points $B$ which were previously filtered are now available for the clustering algorithm. However, the object is only represented by a very limited number of radar points, which hinders the \gls{dbscan} algorithm to form a cluster. At the moment, this means that the detection of very sparsely represented objects still collapses during the clustering step. One way to find these objects is to reduce the number of $min_{pts}$ in the \gls{dbscan} algorithm from $min_{pts} = 4$ to $min_{pts} = 3$ or $min_{pts} = 2$ depending on the criticality of the points.

\subsection{Comparison with the Velocity-based Criticality Metric}
\label{sec:comparison_results}

Additionally, we compared our approach of criticality evaluation to another self-suggesting idea, rating radar points with a high relative velocity as more critical. To keep this experiment at an intelligible level, we suggest to use the ego-motion-compensated doppler velocity $v_{dopp, comp}$. It can be determined with the vehicle's velocity $v_{veh}$ and the angle between the sensor's normal and the radar point:

\begin{align}
	v_{dopp, comp} & = v_{dopp} - v_{veh} \cdot \cos{\varphi}
\end{align}

Points with no relative velocity $v_{dopp, comp}$ receive a criticality value of $crit = 0.0$ and points with a high relative velocity of $|v_{dopp, comp}| \geq \SI{1.5}{\metre\per\second}$ receive a criticality of $crit = 0.0$. Between, the criticality rises linearly: $crit = \frac{|v_{dopp, comp}| - \SI{0.4}{\metre\per\second}}{1.1}$. We selected a filter threshold of \SI{-10.0}{\decibel\metre\squared} and evaluated all scenes of the RadarScenes data set. The posteriori method was applied for both our baseline case as well as the velocity-based criticality metric. The results are shown in table \ref{tab:results_comp}.

\begin{table}[ht]
	\caption{Comparison of point identification and clustering performance between our criticality evaluation method and the velocity-based criticality evaluation method. Superior results are shown in \textbf{bold} font.}
	\centering
	\label{tab:results_comp}
	\begin{tabular}{ccc}
		\toprule
		            & \makecell{Based on drive tube                                        \\ (Ours)} & \makecell{Based on relative velocity\\ (Comparison)}\tabularnewline
		\midrule
		Filter Rate & \SI{10.04}{\percent}          & \SI{14.11}{\percent} \tabularnewline
		\midrule
		\makecell{False Positives                                                          \\ ($A \cap \bar{B}$)} & \textbf{\num{3884178}}                 & \num{8626278}  \tabularnewline
		\midrule
		\makecell{False Negatives                                                          \\ ($\bar{A} \cap B$)} & \textbf{\num{36243}}                  & \num{42193}  \tabularnewline
		\midrule
		\makecell{Total Identified                                                         \\($\sum A$)}   & \num{4086920}                 & \num{9608622}  \tabularnewline
		\midrule
		\makecell{Correctly clustered                                                      \\($B \cap C$)}  & \SI{76.60}{\percent}          & \textbf{\SI{82.04}{\percent}}  \tabularnewline
		\midrule
		\makecell{Total Clustered                                                          \\($\sum C$)}   & \num{5339292}                 & \num{5098049}  \tabularnewline
		\bottomrule
	\end{tabular}
\end{table}

As can be seen, the compared method which is based on the relative velocity of radar points, did perform worse than our method when it comes to identifying falsely deleted points. The number of false negatives increased by \SI{16.4}{\percent} and the number of false positives increased even by \SI{122.1}{\percent}. The reason for the high number of False Positives is that the comparison method found more critical points which were processed \textit{posteriori} after the filter bypass. The total number of clustered radar points is at a comparable level. This shows that the relative number of clustered points is less in a comparison, which suggests that the comparison method included more noise points. However, the higher number of radar points identified as critical points $A$ leads to a higher number of correctly clustered points $B \cap C$. This emphasizes that our post-processing algorithm is also able to work well with other criticality metrics. Walking pedestrians were robustly detected critically, even if they were not part of the drive tube. The finding shows that there is potential to incorporate the relative speed more extensively. Still, we emphasize that the better performance on clustering ground truth critical points $B \cap C$ comes with the expense of a high number of false positives, which leads to significantly higher release rates of planner stopping actions.

%WITHOUT TREATMENT:
%\begin{itemize}
%\item {Used scenes (only scenes with points in Set A)}
%\item {clustered points (with different parameters), Set C}
% \item {[tracked points], Set C'}
%\item {Critical points without trajectory adaption, set B'}
%\item {Critical points with trajectory adaption, set B}
%\item {Points identified as criticals without trajectory adaption, set A'. Probability distribution!}
%\item {Points identified as criticals with trajectory adaption, set A.}
%\item {Probability distribution!}
%\item {False Positives, set $A \cap \bar{B}$}
%\item {False Negatives, set $\bar{A} \cap B$}
%\item {Supercritical points without criticality treatment, set $$B %\cap \bar{C}$$}
%\end{itemize}

%WITH TREATMENT:
%\begin{itemize}
%\item {Successful criticality priori-treatment, set $B \cap C_{priori}$}
%\item {Still non-successful criticality priori-treatment, $B \cap \bar{C}_{priori}$}
%\item {Successful criticality posteriori-treatment, set $B \cap C_{posteriori}$}
%\item {Still non-successful criticality posteriori-treatment, set $B \cap \bar{C}_{posteriori}$}

%\end{itemize}

%Wie würde man das Posteriori Verhalten bewerten? In den nachfolgenden Punktwolken steigt die Clusterhäufigkeit. Man muss hier nachweisen, dass ein critical point aus der $P_t0$ zum Formen eines Clusters/Tracks in den folgenden Punktwolken $P_t1, \dots, P_t5$ führt.

% Wie würde man das Posteriori Verhalten bewerten?
% - entsteht in der $P_t0$ ein Cluster durch das Absenken des Thresholds
% - entsteht durch das zusätzliche Absenken des Thresholds in der $P_t0$ ein Track
\section{Discussion}
\label{sec:discussion}

% Conclusion of results, discussion and limitations (very basic)
The experimental results showed that criticality-aware filtering is a useful method when noise points need to be deleted from point clouds and functional safety shall be guaranteed at the same time. While we were able to reduce the overall point cloud size significantly, it still showed that clusters emerge in critical regions. The number of critical points which are clustered is improved to a constant level, independently of the previously selected filter criteria. Due to lacking labels in the RadarScenes data set, it is still difficult to quantize the impact of the algorithm on the robustness of the object detection. To receive further insights, the pipeline needs to be considered as a whole and radar points must be labeled individually for noise or static object class.

\begin{figure}[ht]
    \centering
    \includegraphics[width = \linewidth]{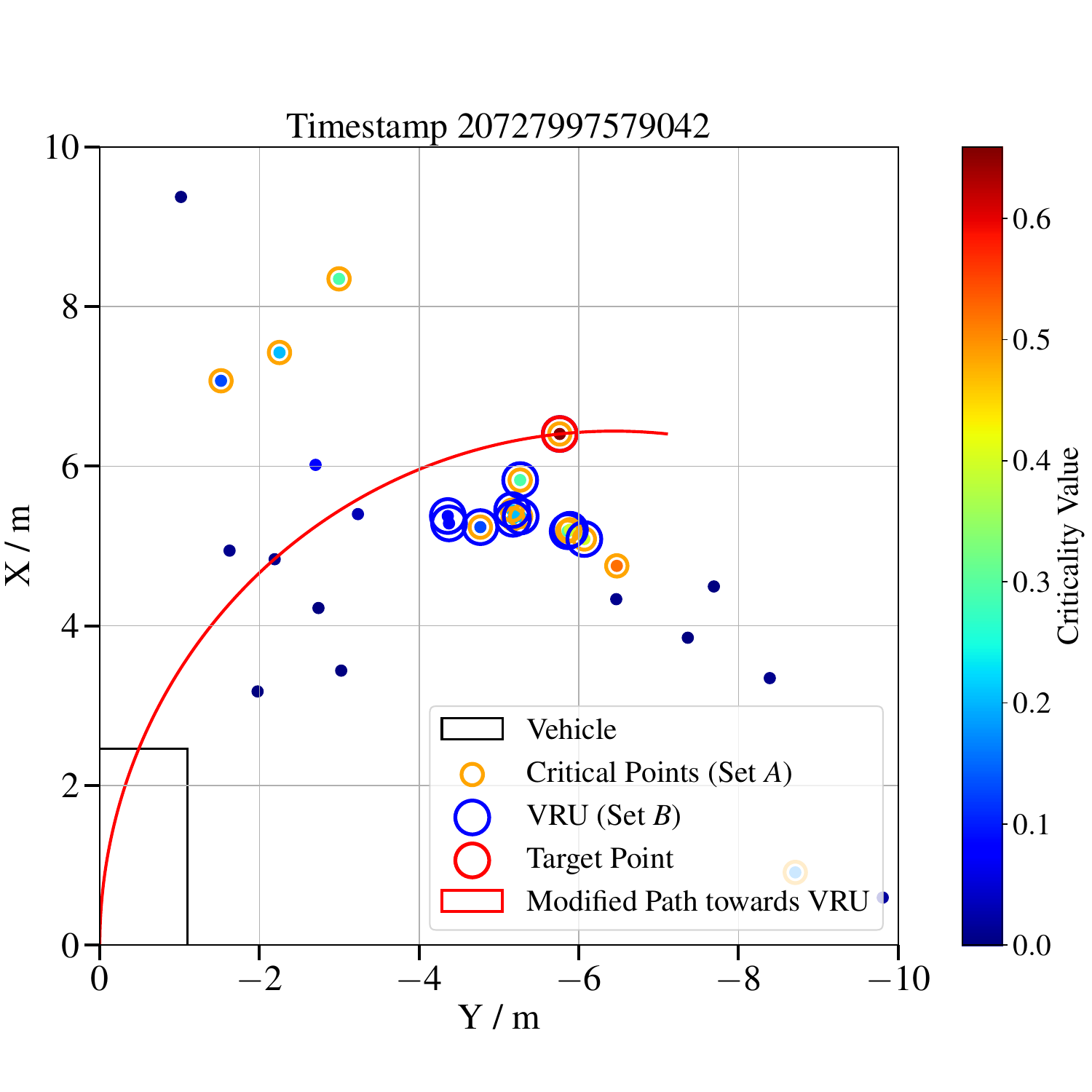}
    \caption{Situation with two reflections of the group $\bar{A} \cap B$, marked by a blue, but no orange circle around}
    \label{fig:limits1}
\end{figure}

% LIMITATION 1: A_bar and B / true critical points which are not classified as critical
However, we could observe limits of the method, which shall be explained by means of the following scenes. A significant hazard which impairs treating all relevant radar points correctly is the upstream radar point criticality assessment. In the experiment, we found that our method received a recall value of \SI{95.02}{\percent}. This means that about \SI{5}{\percent} of all true positive critical points were not detected by our method. Figure \ref{fig:limits1} shows a scene where this happened. In this scene, the path was dragged towards a pedestrian group. All radar points reflected by this group count as true positive critical points and are marked with a blue circle. Most of these points feature an orange circle as well, meaning that they were identified as critical by our algorithm. However, as can be seen, two of these points miss this label. The reason is that these two points are closer to the vehicle and the velocity intended by the planner is too low to consider these points as critical points. It can be reasoned that the used ground truth lacks accuracy in some cases, as all points of one object group are considered critical, independently on their actual position to the vehicle. Objects with higher dimension, e.g. pedestrian groups, may be only partly endangered, if they are not completely located in the drive tube.

% Man findet sie hinter der punktbewertung. Print IF point cloud contains B and A_bar. Kringel = relevante Punkte (B), A_not = rot.

\begin{figure}[ht]
    \centering
    \includegraphics[width = \linewidth]{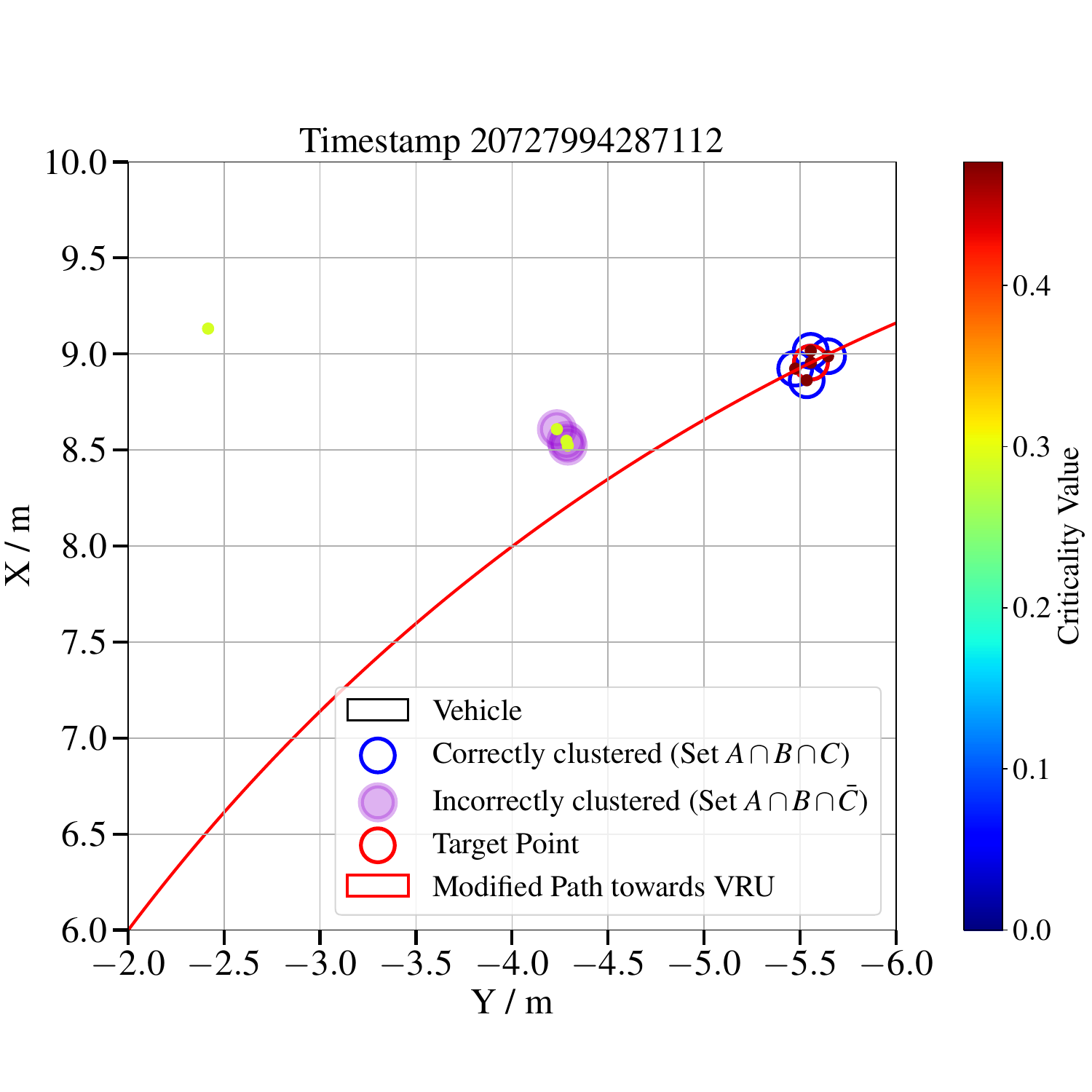}
    \caption{Situation with three reflections of the group $A \cap B \cap \bar{C}$, marked by a violet background. Points being clustered correctly are marked by a blue circle around}
    \label{fig:limits2}
\end{figure}

% LIMITATION 2: A and B and C_bar = blue / points which have not been clustered, though being critical
A second limitation results from the standard clustering algorithm, which is not yet modified for criticality awareness. The radar points are clustered independently from their criticality only if the number of points in this cluster would exceed $min_{Pts} = 4$. Figure \ref{fig:limits2} shows a scene similar to the one before where three radar points of the pedestrian group were not clustered due to their distance to other points of the group ($A \cap B \cap \bar{C}$). These points are marked with a violet background in the figure. A cluster which is not formed may result in a missed object for the planning algorithm. We suggest tackling this problem by forming clusters from less point if these points are evaluated sufficiently critical, as it is the case in this scene. The criticality of each non-clustered point excels $0.3$.

% Man findet sie in der PP nach dem letzten Clustering. Jeder geclusterte Punkt hat eine cluster_ID != -1, jeder kritische Punkt hat point label 5, 6, 8. Kringel = relevante punkte (B), nicht geclustert = rot. cluster_dbscan.cpp hierfuer modifiziert.

% LIMITATION 3: B and F_bar = red / points which have been filtered, though being true critical points

% Man findet sie sehr schwer.

% IMPLICATION: Impact on research and automated driving
In this work, we proved the hypothesis that safety awareness is beneficial in early stages during the perception task. With our method, downstream detection algorithms such as clustering, tracking and classification have a better foundation. Furthermore, we can now guarantee that information about critical objects is not lost due to misinterpreted relevance. This clears the way for the application of sensors in SAE level 4 systems, even if they generally come with a high noise level, such as radar sensors. The methodology is transferable to other sensor types such as ultrasonic sensors. Moreover, with this work, we like to inspire the discussion about a holistic perception task which adapts to the current environmental situation and grants variable attention to objects depending on their criticality.

% LIMITATION: why does clustering not mean that the object is detected correctly
The proposed method is suitable to investigate regions of potential objects in a downstream manner. This means that the awareness time starts in the moment when the criticality region is defined. Conversely, additional three to five measurement cycles are required to create the object track and manifest an object hypothesis. Considering a cycle time of around $\SI{50}{\milli\second}$, the distances which the automated vehicle as well as the \gls{vru} moved during this time can be essential. For this reason, future work will investigate how prevalent, existing measurements can be applied to ensure safety by finding potential clusters earlier in a \textit{priori} processing. Secondly, the impacts of the described delay effect on the emergence of accidents or hazardous situations requires further analysis.

% LIMITATION: no investigation of coherent scenes
Due to the artificial creation of critical scenes, only single measurements could be investigated in this work. The only way to see the impact of our framework on entire sequences is to record an own data set which consists of safety critical situations, labeling critical objects continuously. This would allow to integrate planner actions to draw conclusions about the behavior of the sense-plan pipeline and averting hazardous situations.

% FUTURE WORK: situational awareness, occlusion
In the future, we have various ideas on how to improve the enrichment of critical regions with information. Some of them will be proposed successively. At the moment, we did not integrate any situational awareness into our region identification. Considering that pedestrians are more likely to appear on pedestrian walks or behind vehicles, this information could be used to automatically create criticality regions there. Another benefit is that the detection in occlusion situations is more complicated. These points motivate a prioritized region initialization in such situations.

% FUTURE WORK: move to the front of the processing pipeline
Moreover, the experimental results imply that the modification of the filtering algorithm should happen at a stage as early as possible. Information which is lost on the signal or point level can later not be restored. In contrast to other state-of-the-art approaches, the framework introduced the criticality-awareness on a point level instead of an object list level. However, the thought can be taken further by adapting thresholds on the \gls{dsp} level. With this approach the whole process of forming objects can be controlled, which allows to guarantee error rates.
\section{Conclusion}
\label{sec:conclusion}

In this work, we raised the question how point cloud filtering needs to be tackled when it comes to safety-relevant application fields such as SAE level 4 functions. We believe that a conventional filtering without consideration of safety aspects leads to safety-critical situations if weak points are deleted although they represent a \gls{vru} such as a pedestrian or bicyclist. To address this concern, a metric was elaborated to determine the criticality of each point. Based on this metric, we presented an approach to temporary suspend the filter threshold in regions where critical points have been identified. This procedure guarantees that weak points are not incorrectly deleted in future measurement cycles. The evaluation was performed on the RadarScenes data set, in which artificially generated, critical trajectories towards radar points on \gls{vru}s were added. The labels required for this purpose are included in the data set. These trajectories can be assumed to have hazardous consequences if the planner does not change the intended trajectory. Based on this data set, the criticality metric as well as the safety-aware post-processing method were evaluated. We proved that our method is able to identify true critical radar points robustly and our modified filtering method was able to reduce the number of non-clustered points on \gls{vru}s significantly. The investigations contribute to the applicability of radar sensors in a perception system which acts autonomously without the human driver on a fallback level. The experimental results establish a demand for further research in this field, e.g., for signal processing in the radar pipeline which is upstream of the point cloud.

% \section*{Acknowledgments}
%This publication was written in the context of the RepliCar research project (\href{www.replicar-project.de}{www.replicar-project.de}) funded by the Federal Ministry for Economic Affairs and Climate Action (BMWK). We gratefully acknowledge the financial support.

\bibliographystyle{IEEEtran}
\bibliography{IEEEabrv, ref}

% Generated by IEEEtran.bst, version: 1.14 (2015/08/26)
\begin{thebibliography}{10}
\providecommand{\url}[1]{#1}
\csname url@samestyle\endcsname
\providecommand{\newblock}{\relax}
\providecommand{\bibinfo}[2]{#2}
\providecommand{\BIBentrySTDinterwordspacing}{\spaceskip=0pt\relax}
\providecommand{\BIBentryALTinterwordstretchfactor}{4}
\providecommand{\BIBentryALTinterwordspacing}{\spaceskip=\fontdimen2\font plus
\BIBentryALTinterwordstretchfactor\fontdimen3\font minus
  \fontdimen4\font\relax}
\providecommand{\BIBforeignlanguage}[2]{{%
\expandafter\ifx\csname l@#1\endcsname\relax
\typeout{** WARNING: IEEEtran.bst: No hyphenation pattern has been}%
\typeout{** loaded for the language `#1'. Using the pattern for}%
\typeout{** the default language instead.}%
\else
\language=\csname l@#1\endcsname
\fi
#2}}
\providecommand{\BIBdecl}{\relax}
\BIBdecl

\bibitem{SAE2021}
\BIBentryALTinterwordspacing
{SAE International.}, ``{J3016 - Taxonomy and Definitions for Terms Related to
  Driving Automation Systems for On-Road Motor Vehicles},'' Online, Tech. Rep.,
  April 2021.
\BIBentrySTDinterwordspacing

\bibitem{Yeong2021}
\BIBentryALTinterwordspacing
D.~J. Yeong, G.~Velasco-Hernandez, J.~Barry, and J.~Walsh,
  ``\BIBforeignlanguage{en}{Sensor and {Sensor} {Fusion} {Technology} in
  {Autonomous} {Vehicles}: {A} {Review}},''
  \emph{\BIBforeignlanguage{en}{Sensors}}, vol.~21, no.~6, p. 2140, Jan. 2021.
\BIBentrySTDinterwordspacing

\bibitem{Zhang2023}
\BIBentryALTinterwordspacing
Y.~Zhang, A.~Carballo, H.~Yang, and K.~Takeda, ``Perception and sensing for
  autonomous vehicles under adverse weather conditions: {A} survey,''
  \emph{ISPRS Journal of Photogrammetry and Remote Sensing}, vol. 196, pp.
  146--177, Feb. 2023.
\BIBentrySTDinterwordspacing

\bibitem{Knott1988}
E.~F. Knott, J.~Shaeffer, and M.~T. Tuley, ``Radar cross section,'' in
  \emph{Aspects of modern radar}.\hskip 1em plus 0.5em minus 0.4em\relax Artech
  House Norwood, 1988, pp. 444--448.

\bibitem{Srivastav2023}
\BIBentryALTinterwordspacing
A.~Srivastav and S.~Mandal, ``Radars for {Autonomous} {Driving}: {A} {Review}
  of {Deep} {Learning} {Methods} and {Challenges},'' \emph{IEEE Access},
  vol.~11, pp. 97\,147--97\,168, 2023.
\BIBentrySTDinterwordspacing

\bibitem{Pearce2023}
\BIBentryALTinterwordspacing
A.~Pearce, J.~A. Zhang, R.~Xu, and K.~Wu,
  ``\BIBforeignlanguage{en}{Multi-{Object} {Tracking} with {mmWave} {Radar}:
  {A} {Review}},'' \emph{\BIBforeignlanguage{en}{Electronics}}, vol.~12, no.~2,
  p. 308, Jan. 2023.
\BIBentrySTDinterwordspacing

\bibitem{ester1996density}
M.~Ester, H.-P. Kriegel, J.~Sander, X.~Xu \emph{et~al.}, ``A density-based
  algorithm for discovering clusters in large spatial databases with noise,''
  in \emph{kdd}, vol.~96, no.~34, 1996, pp. 226--231.

\bibitem{Wolf2021}
\BIBentryALTinterwordspacing
M.~Wolf, L.~R. Douat, and M.~Erz, ``Safety-{Aware} {Metric} for {People}
  {Detection},'' in \emph{2021 {IEEE} {International} {Intelligent}
  {Transportation} {Systems} {Conference} ({ITSC})}, Sep. 2021, pp. 2759--2765.
\BIBentrySTDinterwordspacing

\bibitem{Topan2023}
\BIBentryALTinterwordspacing
S.~Topan, Y.~Chen, E.~Schmerling, K.~Leung, J.~Nilsson, M.~Cox, and M.~Pavone,
  ``Refining {Obstacle} {Perception} {Safety} {Zones} via {Maneuver}-{Based}
  {Decomposition},'' in \emph{2023 {IEEE} {Intelligent} {Vehicles} {Symposium}
  ({IV})}, Jun. 2023, pp. 1--8, iSSN: 2642-7214.
\BIBentrySTDinterwordspacing

\bibitem{schonemann2019maneuver}
V.~Sch{\"o}nemann, M.~Duschek, and H.~Winner, ``Maneuver-based adaptive safety
  zone for infrastructure-supported automated valet parking.'' 2019.

\bibitem{Liu2023}
\BIBentryALTinterwordspacing
S.~Liu, L.~Sha, and T.~Abdelzaher,
  ``\BIBforeignlanguage{en}{Criticality-{Based} {Data} {Segmentation} and
  {Resource} {Allocation} in {Machine} {Inference} {Pipelines}},'' M.~Srivatsa,
  T.~Abdelzaher, and T.~He, Eds.\hskip 1em plus 0.5em minus 0.4em\relax Cham:
  Springer International Publishing, 2023, pp. 335--352.
\BIBentrySTDinterwordspacing

\bibitem{Lyssenko2022}
\BIBentryALTinterwordspacing
M.~Lyssenko, C.~Gladisch, C.~Heinzemann, M.~Woehrle, and R.~Triebel, ``Towards
  {Safety}-{Aware} {Pedestrian} {Detection} in {Autonomous} {Systems},'' in
  \emph{2022 {IEEE}/{RSJ} {International} {Conference} on {Intelligent}
  {Robots} and {Systems} ({IROS})}, Oct. 2022, pp. 293--300, iSSN: 2153-0866.
\BIBentrySTDinterwordspacing

\bibitem{Lin_2017_ICCV}
T.-Y. Lin, P.~Goyal, R.~Girshick, K.~He, and P.~Dollar, ``Focal loss for dense
  object detection,'' in \emph{Proceedings of the IEEE International Conference
  on Computer Vision (ICCV)}, Oct 2017.

\bibitem{Philipp2022}
\BIBentryALTinterwordspacing
R.~Philipp, J.~Rehbein, F.~Grün, L.~Hartjen, Z.~Zhu, F.~Schuldt, and F.~Howar,
  ``Systematization of {Relevant} {Road} {Users} for the {Evaluation} of
  {Autonomous} {Vehicle} {Perception},'' in \emph{2022 {IEEE} {International}
  {Systems} {Conference} ({SysCon})}, Apr. 2022, pp. 1--8, iSSN: 2472-9647.
\BIBentrySTDinterwordspacing

\bibitem{antonante2023task}
P.~Antonante, S.~Veer, K.~Leung, X.~Weng, L.~Carlone, and M.~Pavone,
  ``Task-aware risk estimation of perception failures for autonomous vehicle,''
  in \emph{Robotics: Science and Systems (RSS)}.\hskip 1em plus 0.5em minus
  0.4em\relax Robotics: Science and Systems (RSS), 2023.

\bibitem{Ceccarelli2024}
\BIBentryALTinterwordspacing
A.~Ceccarelli and L.~Montecchi, ``Object criticality for safer navigation,''
  Tech. Rep., Apr. 2024, arXiv:2406.10232 [cs] type: article.
\BIBentrySTDinterwordspacing

\bibitem{Feng2024}
\BIBentryALTinterwordspacing
Y.~Feng and Y.~Sun, ``{PolarPoint}-{BEV}: {Bird}-eye-view {Perception} in
  {Polar} {Points} for {Explainable} {End}-to-end {Autonomous} {Driving},''
  \emph{IEEE Transactions on Intelligent Vehicles}, pp. 1--11, 2024.
\BIBentrySTDinterwordspacing

\bibitem{Major2019}
B.~Major, D.~Fontijne, A.~Ansari, R.~Teja~Sukhavasi, R.~Gowaikar, M.~Hamilton,
  S.~Lee, S.~Grzechnik, and S.~Subramanian, ``Vehicle detection with automotive
  radar using deep learning on range-azimuth-doppler tensors,'' in
  \emph{Proceedings of the IEEE/CVF International Conference on Computer Vision
  (ICCV) Workshops}, Oct 2019.

\bibitem{Matsunami2012}
\BIBentryALTinterwordspacing
I.~Matsunami, R.~Nakamura, and A.~Kajiwara, ``{RCS} measurements for vehicles
  and pedestrian at 26 and {79GHz},'' in \emph{2012 6th {International}
  {Conference} on {Signal} {Processing} and {Communication} {Systems}}, Dec.
  2012, pp. 1--4.
\BIBentrySTDinterwordspacing

\bibitem{Chen2014}
\BIBentryALTinterwordspacing
M.~Chen and C.-C. Chen, ``{RCS} {Patterns} of {Pedestrians} at 76-77 {GHz},''
  \emph{IEEE Antennas and Propagation Magazine}, vol.~56, no.~4, pp. 252--263,
  Aug. 2014.
\BIBentrySTDinterwordspacing

\bibitem{Schumann2021}
\BIBentryALTinterwordspacing
O.~Schumann, M.~Hahn, N.~Scheiner, F.~Weishaupt, J.~F. Tilly, J.~Dickmann, and
  C.~Wöhler, ``{RadarScenes}: {A} {Real}-{World} {Radar} {Point} {Cloud}
  {Data} {Set} for {Automotive} {Applications},'' in \emph{2021 {IEEE} 24th
  {International} {Conference} on {Information} {Fusion} ({FUSION})}, Nov.
  2021, pp. 1--8.
\BIBentrySTDinterwordspacing

\end{thebibliography}

% \vspace{5pt}

\begin{IEEEbiographynophoto}{Tim Brühl}
received the M. Sc. degree in Electrical Engineering and Information Technology from Karlsruhe Institute of Technology in 2022. He is currently a doctoral candidate at Karlsruhe Institute of Technology and a researcher at Dr. Ing. h.c. F. Porsche AG. His research interests include safety aspects of radar point cloud processing, clustering, tracking and radar-camera fusion in the application field of automated parking functions.
\end{IEEEbiographynophoto}
\begin{IEEEbiographynophoto}{Jenny Glönkler}
received the M. Eng. degree in Automotive Engineering with focus on Advanced Driver Assistance Systems and Autonomous Driving from Esslingen University of Applied Sciences in 2024. During her master's thesis at Dr. Ing. h.c. F. Porsche AG, she worked on safety aspects of radar and camera-based perception systems, particularly the identification and handling of SOTIF-relevant scenarios for automated parking functions according to SAE Level 4.
\end{IEEEbiographynophoto}
\begin{IEEEbiographynophoto}{Robin Schwager}
received the M.Sc. degree in Mechatronic System Engineering from the
Pforzheim University of Applied Sciences. He is currently a doctoral candidate at the Karlsruher Institute of Technology and Dr. Ing. h.c. F. Porsche AG. His research interests include the field of individualization and optimization of automated and assisted driving functions.
\end{IEEEbiographynophoto}
\begin{IEEEbiographynophoto}{Tin Stribor Sohn}
received the M.Sc. degree in Computer Science from the University of Tuebingen. He is currently a doctoral candidate at Karlsruhe Institute of Technology and researcher at Dr. Ing. h.c. F. Porsche AG. His research interests include the data-driven identification of triggering conditions, computer vision and scenario-based testing of automated driving systems.
\end{IEEEbiographynophoto}
\begin{IEEEbiographynophoto}{Tim Dieter Eberhardt}
received the M.Sc. degree in Mechanical Engineering from HTW Berlin. He is currently a doctoral candidate at the Karlsruhe Institute of Technology, conducting research in collaboration with Dr. Ing. h.c. F. Porsche AG. His research focuses on Advanced Driver Assistance Systems (ADAS), with a particular emphasis on Machine Learning, Computer Vision, and Autonomous Systems. His academic and professional work centers around the integration of new technologies to enhance vehicle safety.
\end{IEEEbiographynophoto}
\begin{IEEEbiographynophoto}{Sören Hohmann}
studied Electrical Engineering at the Technische Universität Braunschweig, University of Karlsruhe and école nationale supérieure d’électricité et de mécanique Nancy. He received his diploma degree and Ph.D. degree at the University of Karlsruhe. Afterwards, he worked for BMW in different executive positions. Since 2010, he is the head of the IRS at the KIT. In addition, he is a director of the FZI Research Center for Information Technology. His research interests are cooperative control systems, control of energy systems and functional safety control. He is senior member of the IEEE and chair of the IFAC technical committee on human machine systems.
\end{IEEEbiographynophoto}

\vfill

\end{document}